%% file: main.tex
\newif\ifpreprint
\newif\ifauthorversion

\preprinttrue
\ifpreprint
\documentclass[sigconf,screen,nonacm]{acmart}
\else
\ifauthorversion
\documentclass[sigconf,screen,authorversion]{acmart}
\else
\documentclass[sigconf]{acmart}
\fi
\fi

\AtBeginDocument{%
  \providecommand\BibTeX{{%
    \normalfont B\kern-0.5em{\scshape i\kern-0.25em b}\kern-0.8em\TeX}}}

\copyrightyear{2021}
\acmYear{2021}
\setcopyright{rightsretained}
\acmConference[CHI '21]{CHI Conference on Human Factors in Computing Systems}{May 8--13, 2021}{Yokohama, Japan}
\acmBooktitle{CHI Conference on Human Factors in Computing Systems (CHI '21), 
May 8--13, 2021, Yokohama, Japan}\acmDOI{10.1145/3411764.3445648} 
\acmISBN{978-1-4503-8096-6/21/05}

\begin{document}

% File of custom macros
\input{macros}

% Other customizations
\def\UrlBigBreaks{\do\/\do-\do\#}

\def\plaintitle{Augmenting Scientific Papers with Just-in-Time, 
Position-Sensitive Definitions of Terms and Symbols}

\title[Augmenting Scientific Papers with Definitions of Terms and 
Symbols]{\plaintitle}

\settopmatter{authorsperrow=4}
\author{Andrew Head}
\email{andrewhead@berkeley.edu}
\orcid{0000-0002-1523-3347}
\affiliation{UC Berkeley}
\author{Kyle Lo}
\email{kylel@allenai.org}
\affiliation{Allen Institute for AI}
\orcid{0000-0002-1804-2853}
\author{Dongyeop Kang}
\email{dongyeopk@berkeley.edu}
\affiliation{UC Berkeley}
\author{Raymond Fok}
\email{rayfok@cs.washington.edu}
\affiliation{University of Washington}
\author{Sam Skjonsberg}
\email{sams@allenai.org}
\affiliation{Allen Institute for AI}
\author{Daniel S. Weld}
\email{danw@allenai.org}
\affiliation{Allen Institute for AI}
\affiliation{University of Washington}
%\additionalaffiliation{University of Washington}
\author{Marti A. Hearst}
\email{hearst@berkeley.edu}
\affiliation{UC Berkeley}
\orcid{0000-0002-4346-1603}

\renewcommand{\shortauthors}{Head et al.}

\begin{abstract}
\input{00-abstract}
\end{abstract}

\ifpreprint
\else
% Regenerate at https://dl.acm.org/ccs/ccs_flat.cfm
\fi

% has been: analytic reading, scientific papers, literature graphs, cross-linked 
% navigation, documents as code, simple vision, page previews.
\keywords{interactive documents, \ifpreprint\else reading interfaces,\fi scientific papers, 
definitions, nonce words}


\maketitle

\input{01-introduction}
\input{02-prior-work}
\input{03-reading-study}
\input{04-design-motivations}
\input{05-system}
\input{06-implementation}
\input{07-evaluation}
\input{08-agenda}
\input{09-conclusion}

% Every figure should also have a figure description unless it is purely
% decorative. These descriptions convey what’s in the image to someone
% who cannot see it. They are also used by search engine crawlers for
% indexing images, and when images cannot be loaded.
% 
% A figure description must be unformatted plain text less than 2000
% characters long (including spaces).  {\bfseries Figure descriptions
%   should not repeat the figure caption – their purpose is to capture
%   important information that is not already provided in the caption or
%   the main text of the paper.} For figures that convey important and
% complex new information, a short text description may not be
% adequate. More complex alternative descriptions can be placed in an
% appendix and referenced in a short figure description. For example,
% provide a data table capturing the information in a bar chart, or a
% structured list representing a graph.  For additional information
% regarding how best to write figure descriptions and why doing this is
% so important, please see
% \url{https://www.acm.org/publications/taps/describing-figures/}.

\begin{acks}
Zachary Kirby, Jocelyn Sun, Luming Chen, Nidhi Kakulawaram, RJ Pimentel, and 
Benjamin Barantschik contributed to the design, implementation, and evaluation 
of prototypes of ScholarPhi. Luca Weihs, Brendan Roof, and Alvaro Herrasti 
developed a prototype algorithm for locating equations in \LaTeX{} papers which 
inspired the approach used to locate symbols and terms in this paper. Luca Weihs and Amrit Dhar provided feedback on the statistical analysis. This work 
would not have been possible without their contributions.

This research receives funding from the Alfred P.  Sloan Foundation, the Allen 
Institute for AI,  Office of Naval 
Research  grants \texttt{N00014-15-1-2774} and \texttt{N00014-18-1-2193}, National Science Foundation RAPID grant 2040196, and the Washington Research Foundation Thomas J. Cable Professorship.
\end{acks}

\bibliographystyle{ACM-Reference-Format}
\bibliography{cleaned-references}

%\appendix

\input{10-appendix}

\end{document}

%% file: macros.tex
\def\systemname{%
ScholarPhi%
}

\newcommand{\andrew}[1]{\textcolor{blue}{[#1 -A]}}
\newcommand{\kyle}[1]{\textcolor{red}{[#1 -K]}}
\newcommand{\sam}[1]{\textcolor{brown}{[#1 -S]}}
\newcommand{\marti}[1]{\textcolor{green}{[#1 -M]}}
\newcommand{\dan}[1]{\textcolor{purple}{[#1 -D]}}
\newcommand{\dongyeop}[1]{\textcolor{cyan}{[#1 -D]}}
\newcommand{\jocelyn}[1]{\textcolor{gold}{[#1 -D]}}
\newcommand{\ray}[1]{\textcolor{orange}{[#1 -R]}}

% Documenting changes
% \newcommand{\change}[1]{\textcolor{purple}{#1}}
% \newenvironment{changes}
% {\begingroup\color{purple}}
% {\endgroup}

% When submitting, swap out the above with the following. Do this
% instead of disabling colors altogether, as the color commands can
% introduce subtle spacing differences in the paper.
\newcommand{\change}[1]{\textcolor{black}{#1}}
\newenvironment{changes}
{\begingroup\color{black}}
{\endgroup}

\newcommand{\plain}{{\textsc{Basic}}}
\newcommand{\sesame}{{\textsc{Declutter}}}
\newcommand{\everything}{{\textsc{ScholarPhi}}}

\newcommand{\qone}{{\textsc{Results}}}
\newcommand{\qtwo}{{\textsc{Dataset}}}
\newcommand{\qthree}{{\textsc{Symbols}}}

\newcommand{\acc}{{\textsc{Correct}}}
\newcommand{\ease}{{\textsc{Ease}}}
\newcommand{\conf}{{\textsc{Confidence}}}
\newcommand{\timeVar}{{\textsc{Time}}}
\newcommand{\dist}{{\textsc{Distance}}}
\newcommand{\area}{{\textsc{Area}}}

% Inline figures
\newenvironment{inlinefigureenv}
{\setlength{\topsep}{2.5ex}\center}
{\endcenter}

\newcommand{\inlinefigure}[2][.45\textwidth]{%
\begin{inlinefigureenv}%
\includegraphics[width=#1]{#2}%
\vspace{-1.25ex}%
\end{inlinefigureenv}%
}

\def\KaTeX{K\kern-.2em\raisebox{.2em}{\scriptsize A}\kern-.12em\TeX}

% Dotted underline
% \newcommand{\udensdot}[1]{%
%     \tikz[baseline=(todotted.base)]{
%         \node[inner sep=1pt,outer sep=0pt] (todotted) {#1};
%         \draw[gray, densely dotted] (todotted.south west) -- (todotted.south east);
%     }%
% }%

%% file: 00-abstract.tex
Despite the central importance of research papers to scientific progress, they 
can be difficult to read. Comprehension is often stymied when the information 
needed to understand a passage resides somewhere else---in another section, or 
in another paper. In this work, we envision how interfaces can bring definitions 
of technical terms and symbols to readers when and where they need them most. We 
introduce \emph{ScholarPhi}, an augmented reading interface with four novel 
features: (1) tooltips that surface position-sensitive definitions from 
elsewhere in a paper,  (2) a filter over the paper that ``declutters'' it to 
reveal how the term or symbol is used across the paper, (3) automatic equation 
diagrams that expose multiple definitions in parallel, and (4) an automatically 
generated glossary of important terms and symbols. A usability study showed that 
the tool helps researchers of all experience levels read papers.  Furthermore, 
researchers were eager to have ScholarPhi's definitions available to support 
their everyday reading.

%% file: 01-introduction.tex
\section{Introduction}

\begin{figure}
\centering
\includegraphics[width=.47\textwidth]{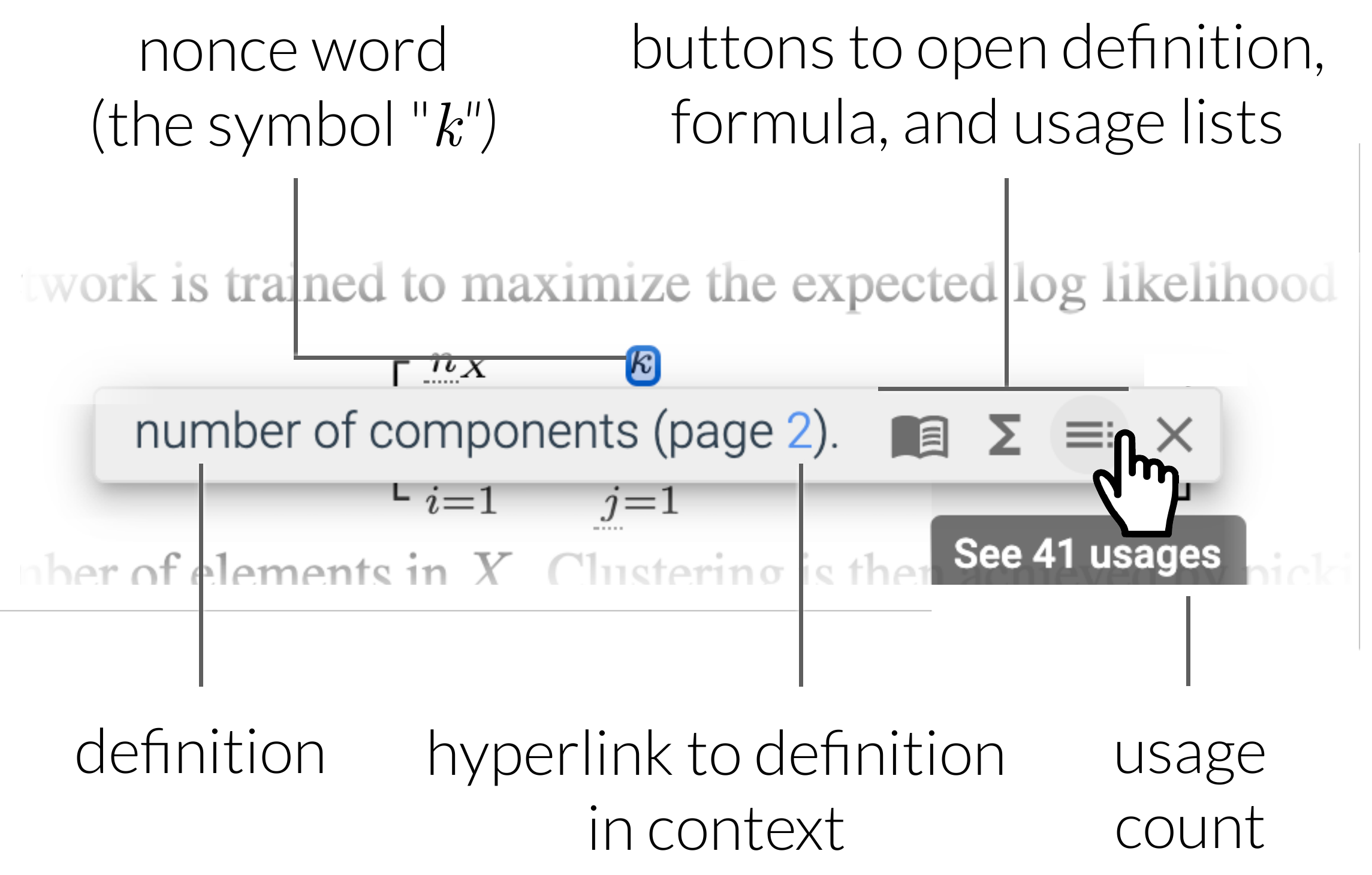}
\caption{%
ScholarPhi helps readers understand \emph{nonce words}---unique technical terms 
and symbols---defined within scientific papers. \textmd{When a reader comes 
across a nonce word that they do not understand, ScholarPhi lets them click the 
word to view a position-sensitive definition in a compact tooltip. The tooltip 
lets the reader jump to the definition in context. It also lets them open lists 
of prose definitions, defining formulae, and usages of the word.  ScholarPhi 
augments the reading experience with this and a host of other features (see 
Section~\ref{sec:demo}) to assist readers.}}
\label{fig:teaser}
\Description{%
Screenshot of a tooltip that defines a nonce word. The tooltip appears beneath 
the symbol ``k''.

The width of the tooltip is about half the width of the paper. The height of the 
tooltip is about the same as the height of one line of text.

The definition of ``k'' is ``number of components.'' To the right of the 
definition are the words ``page 2,'' which link to where the definition can be 
found.

The tooltip also contains buttons for opening lists of all textual definitions 
of ``k'' anywhere in the paper, all formulae that define the symbol, and all 
usages of the symbol. There is also a button to dismiss the tooltip.
}
\end{figure}

Researchers are charged with keeping on top of immense, rapidly-changing 
literatures. Naturally, then, reading constitutes a major part of a researcher's
everyday work. Senior researchers, such as faculty members,
spend over one hundred hours a 
year reading the literature, consuming over one hundred papers annually~\cite{ref:tenopir2009electronic}. And despite the formidable background knowledge that a researcher
gains over the course of their career, they will still often find that papers
are prohibitively difficult to read.

\begin{figure}

\centering
\includegraphics[width=.45\textwidth]{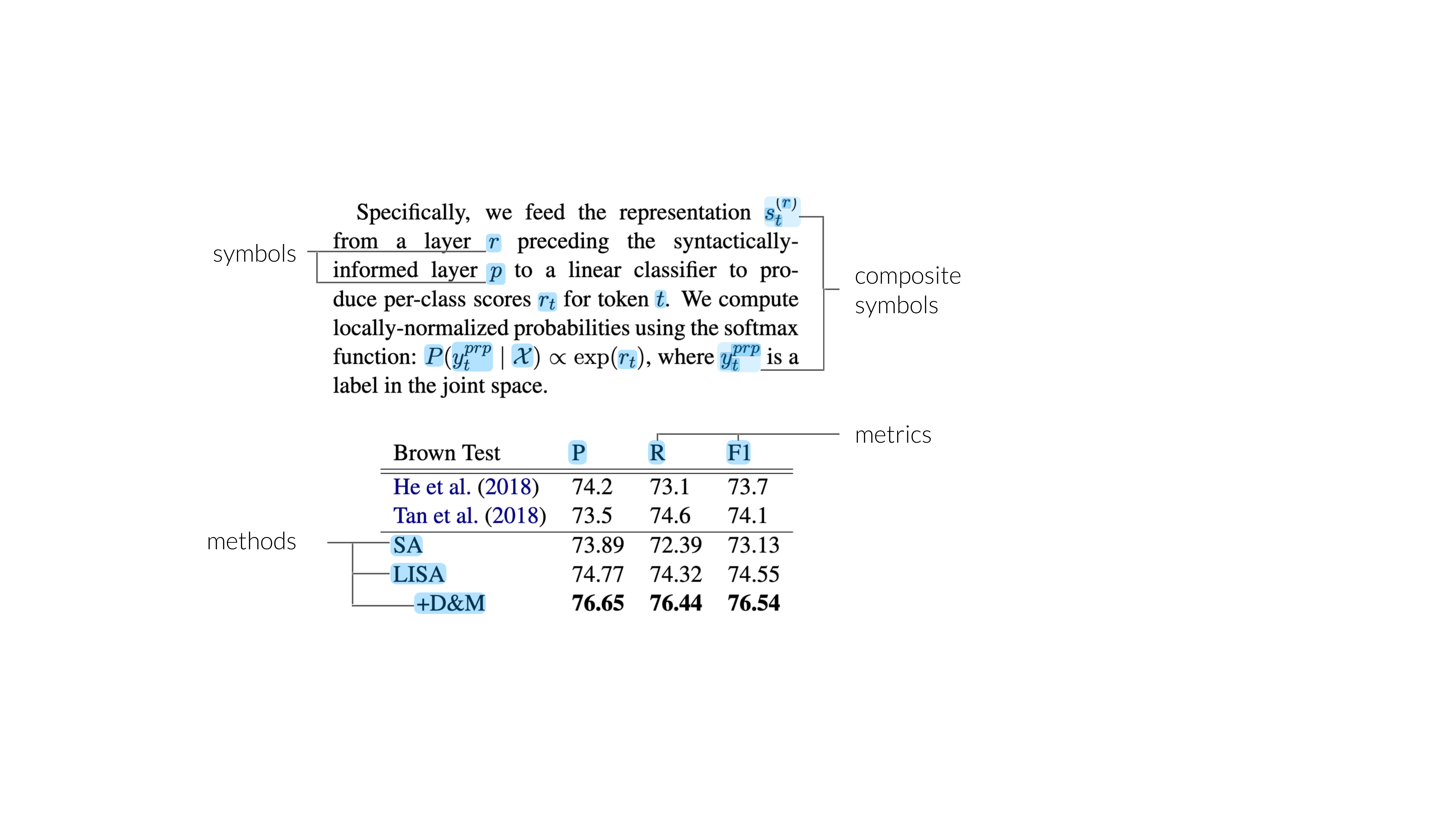}
\caption{One challenge to reading a paper is making sense of the hundreds of 
nonce words within them. \textmd{Nonce words, like the symbols, abbreviations, 
and terms shown above, are defined within a paper for use within that paper. As 
such, a reader cannot know what they mean ahead of time.  Quintessential 
examples of nonce words in the computer science literature are mathematical 
symbols, and abbreviations for metrics, algorithms, and datasets. The excerpts above
are from Strubell et al.'s \emph{Linguistically-informed self-attention for
semantic role labeling}~\cite{ref:strubell2018linguistically}.}}
\label{fig:abundance_of_nonces}
\Description{%
Two passages from a paper with nonce words highlighted.

The first passage is a paragraph from the algorithm section of the paper. It 
contains 10 symbols. Some symbols are simple, comprising single characters 
(e.g., ``r,'' ``p''). Others are composite, made up of many characters, like 
``r-sub-t.''

The second passage is a table of results. It contains six abbreviations local to 
this paper. Three abbreviations refer to methods evaluated in the paper, and 
three to metrics like precision and recall.
}
\end{figure}

As they read, a researcher is constantly trying to fit the information they find 
into schemas of their prior knowledge, but the success of this assimilation is 
by no means guaranteed~\cite{ref:bazerman1985physicists}. A researcher may
struggle to understand a paper due to gaps in their own knowledge, or due to the
intrinsic difficulty of reading a specific 
paper~\cite{ref:bazerman1985physicists}. Reading is made all the more 
challenging by the fact that scholars increasingly read selectively,  looking 
for specific information by skimming and 
scanning~\cite{ref:nicholas2010researchers,ref:hillesund2010digital,ref:late2019changes}.

We are motivated by the question: Can a novel  interface improve the reading 
experience by reducing  distractions that interrupt the reading 
flow?  This work takes a measured step to address the general design question 
by focusing on the specific case of helping readers understand cryptic technical 
terms and symbols defined within a paper, which are called ``nonce words''  in 
the field of linguistics.  Formally, a \emph{nonce word} is a word that is
coined for a particular use, which is unlikely to become a permanent part of the 
vocabulary~\cite{ref:mattiello2017analogy}. Because a nonce word is localized to a 
specific paper, a reader cannot know precisely what it means when they start 
reading the paper. Because it is only intended for use within a single paper, it 
is likely to be defined somewhere within that same paper, but finding that 
definition may require significant effort by the reader.  By their nature, nonce 
words are an interesting focus for augmenting reading tools because readers will 
have questions about them, and those questions will be answerable (exclusively 
by) searching the text that contains them.

Two aspects of nonce words constrain the design of any reading application that 
is built to define them. First, they are numerous: a paper can contain hundreds 
of them. Indeed, a single passage or table may contain a dozen terms closely 
packed together
(see Figure~\ref{fig:abundance_of_nonces}). In such settings 
a reader is likely to have demands on their working memory and may also want to 
see definitions for multiple nonce words in the same vicinity.

Second, nonce words are sometimes assigned multiple definitions within the same 
paper.  One example is a symbol like $k$, which over the course of a single 
paper may variously stand for a dummy variable in a summation operation, the 
number of components in a mixture of Gaussian models, and the number of clusters 
output by a clustering algorithm (see the scenario in Section~\ref{sec:demo}).  
These two aspects of nonce words raise the question of whether conventional 
solutions for showing definitions of terms (e.g., the electronic glossaries 
explored in second-language learning 
research~\cite{ref:cheng2009l1glosses,ref:yanagisawa2020different} or 
Wikipedia's page previews~\cite{ref:wikipediapagepreviews}) also suit a
researcher who is puzzling their way through dense, cryptic, ambiguous
notation.

\change{In this work, we introduce \emph{ScholarPhi}, a tool that helps readers
efficiently access definitions of nonce words in scientific papers. The larger vision of
ScholarPhi is to help scientists more easily read papers by linking
relevant information to its location of use. We envision the tool eventually providing
access to the contents of cited papers, and definitions external to the paper. The current paper focuses on one portion of this problem: the design and evaluation of interfaces for understanding nonce words.}

This paper begins with a formative study of nine readers as they read a 
scientific text of their own choice (Section~\ref{sec:formative}). Most readers 
expressed confusion at nonce words in their texts. Many readers were reluctant 
to look up what the words meant, given the anticipated cost of doing so. This 
inspired the subsequent design of a tool that could have answered those readers' 
questions while reducing friction so that readers would actually use the tool.

We then describe design motivations for a new reading interface 
(Section~\ref{sec:design-motivations}) that are grounded in insights from four 
pilot studies with early prototypes, conducted with 24 researchers.
Key insights from the research include the importance of tailoring definitions 
to the passage where a reader seeks to understand a nonce word, and the 
competing goals of providing scent (i.e., visual 
cues~\cite{ref:pirolli1999information}) of what is defined without distracting 
from a reading task that is already cognitively demanding on its own.

Building on the motivations found in the pilot research, a user interface is
presented (Section~\ref{sec:demo}). The basic design of ScholarPhi is one of an 
interactive hypertext interface. A reader's paper is augmented with subtle 
hyperlinks indicating which nonce words can be clicked in order to access 
definition information. Readers can click nonce words to access definitions for 
those words in a compact tooltip (Figure~\ref{fig:teaser}). These definitions 
are position-sensitive---that is, if there are multiple definitions of a nonce 
word in the text, ScholarPhi uses the heuristic of showing readers the most 
recent definition that appears before the selected usage of the word.  
Definitions are also linked to the passage they were extracted from: a reader 
can click on a hyperlink next to the definition to jump to where it appears in 
the paper.  In addition to definitions, the tooltip makes available a list of 
all usages of the nonce word throughout the text, as well as a special view of 
formulae that include the word.

Beyond these basic affordances, ScholarPhi provides a suite of features, each of 
which provides readers with efficient yet non-intrusive methods for accessing 
information about nonce words. First, ScholarPhi provides efficient, precise 
selection mechanics for selecting mathematical symbols and their sub-symbols 
through single clicks, rather than error-prone text selections (Section~\ref{sec:definition-tooltips}). Second, 
ScholarPhi provides a novel filter over the paper called ``declutter'' that 
helps a reader search for information about a nonce word by low-lighting all 
sentences in the paper that do not include that word 
(Section~\ref{sec:declutter}). Third, ScholarPhi generates equation diagrams and 
overlays them on top of display equations, affixing labels to all symbols and 
sub-symbols in the equation for which definitions are available 
(Section~\ref{sec:equation-diagrams}). The final feature is a priming glossary 
comprising definitions of all nonce words that appear in a paper, prepended to 
the start of the document (Section~\ref{sec:priming-glossary}).

The emphasis in the design of each of these features is on acknowledging the 
inherent complexity of the setting of scientific papers, and hence designing 
features for looking up definitions that are easy to invoke and minimally 
distracting.

To enable these features, new methods were introduced for analyzing 
scientific papers in order to make nonce words interactive.  A paper processing 
pipeline was built that automatically segments equations into symbols and their 
sub-symbols, detects all usages for a nonce word, and detects precise 
bounding box locations of nonce words so that they may be clicked. \change{The 
implementation of the pipeline is described in Section 
~\ref{sec:current-implementation}. The suitability of contemporary definition 
extraction algorithms is discussed, highlighting a need for improvements to
technologies for definition extraction.}

This work concludes with a controlled usability study with twenty-seven 
researchers (Section~\ref{sec:evaluation}). Researchers were observed as they 
used three versions of  ScholarPhi---one with all the features described 
above, one with only the ``declutter'' feature, and one that behaved exactly 
like a standard, un-augmented PDF reader.

When readers had access to ScholarPhi's features, they could answer questions 
about a scientific paper in significantly less time, while viewing significantly 
less of the paper in order to come to an answer. They reported that they found 
it easier to answer questions about the paper, and were more confident about 
their answers with ScholarPhi. Researchers were also observed as they used 
ScholarPhi for 15 minutes of unstructured reading time.  Researchers made use of 
all of ScholarPhi's features. Feedback was overwhelmingly positive.
Most participants expressed interest in using the features ``often'' or 
``always'' for future papers, with an emphasis on the usefulness of definition 
tooltips and equation diagrams.

In summary, this work makes four contributions. First, it characterizes the 
problem of searching for information about nonce words as one of the challenges of 
reading scientific papers, grounded in a small formative study.  Second, it 
provides design motivations for designing interactive tools that define nonce 
words, grounded in iterative evaluations of prototypes of a tool.  Third, it 
presents ScholarPhi, an augmented reading interface with a suite of novel 
features for helping readers understand nonce words in scientific papers.  
Finally, it provides evidence of the usefulness of the design in searching and 
reading scientific papers through a controlled study with twenty-seven 
researchers.\footnote{An interactive demo, video figure,  and source code for this work can be found on the project website at \url{https://scholarphi.org}.}

%% file: 02-prior-work.tex
\section{Background and Related Work}
\label{sec:related-work}

\subsection{How Researchers Read Papers}

Researchers read papers to become aware of foundational ideas and to stay 
apprised of the latest developments in their field. However, reading papers is 
difficult.  Challenges in reading a paper can come from gaps in a reader's 
knowledge, or from ideas in the paper that are poorly 
explained~\cite{ref:bazerman1985physicists}. Papers may be read out-of-order and 
piecemeal~\cite{ref:bazerman1985physicists,ref:hillesund2010digital,ref:nicholas2010researchers}.  
As a result, a passage of a paper may be read out of context. An additional
challenge is assimilating information scattered across one or many
documents, a challenge common to the activity of active reading in many domains~\cite{ref:ohara1996towards,ref:adler1998diary,ref:tashman2011active}.

Papers that include mathematical content can impose additional demands on a 
reader. Reading mathematical texts often entails grappling with unfamiliar 
terminology and notational idioms, which can be particularly challenging for 
less experienced readers~\cite{ref:shepherd2014reading}. Self-reports from 
mathematicians have suggested that the process of reading math involves 
backtracking as a reader attempts to scaffold their 
understanding~\cite{ref:weber2008how}, a pattern which has also been 
observed in eye-tracking studies of reading 
math~\cite{ref:inglis2012expert,ref:kohlhase2018discourse}. When attempting to 
understand an equation, readers will look to nearby equations and text for 
clarifications~\cite{ref:kohlhase2018discourse}.

While reading papers in physical volumes and print-outs used to be the norm, it 
is increasingly the case that researchers consult papers in digital reading 
applications~\cite{ref:tenopir2009electronic,ref:late2019reading}, particularly 
for some types of scholarly communication such as conference 
proceedings~\cite{ref:late2019reading}. This suggests the value of investing in 
reading user interfaces that take advantage of the 
unique interactive potential of digital interfaces to augment the reading 
experience.

\subsection{Augmented Reading Interfaces}

Since the beginning of human-computer interaction as a discipline, one of the  
foundational challenges has been equipping knowledge workers with tools that  
extend their cognition during reading. Vannevar Bush, in his vision of the memex,  
proposed a system that enabled readers to build trails across the literature,  
linking passages across related readings in a way that made implicit 
connections clear~\cite{ref:bush1945we}. This vision has expressed itself in 
many  forms, from the invention of hypertext~\cite{ref:conklin1987hypertext} 
to experiments with  interactive books~\cite{ref:landauer1993enhancing,ref:norman2013design}  and ``fluid  
documents'' that can adapt their form and content to elaborate where readers  
need clarification~\cite{ref:chang1998negotiation}. In the first decade of the CHI 
conference, myriad techniques were proposed to help readers navigate text using 
social annotations~\cite{ref:hill1992edit}, to augment hypertexts with glosses that 
could dynamically change the layout of the 
text~\cite{ref:zellweger1998fluid}, and to provide navigational affordances that 
allow readers to see overviews of document content and jump quickly to passages of 
interest~\cite{ref:graham1999reader,ref:schilit1998beyond}.

\subsubsection{Glossaries, Definitions, and Explanations}

Today, many reading 
and editing tools show dictionary definitions when a reader hovers over or clicks on a word.
The Word Wise feature in the Amazon 
Kindle lets readers view definitions of tricky words in the space between consecutive lines of text~\cite{ref:kindlewordwise}. In 2014, Wikipedia began to roll out page previews as a feature that allowed readers to preview the content of a referenced page by hovering over a link to that page. Based on positive 
usability evaluation results, Wikipedia decided to make the feature a permanent 
fixture on the site~\cite{ref:wikipediapagepreviews}. Recent proceedings of 
human-computer interaction conferences have introduced prototypes that allow 
readers to answer their questions about how to use web 
pages~\cite{ref:chilana2012lemonaid}, the meaning of cryptic programming 
syntax~\cite{ref:head2015tutorons}, hard-to-visualize 
quantities~\cite{ref:hullman2018improving}, and unfamiliar words from a second language~\cite{ref:lungu2018aswe}.

\subsubsection{Symbol Selection}

ScholarPhi uses an advanced symbol selection technique that draws from related work.
\citet{ref:zeleznik2010hands} 
introduced gestures for a multi-touch display that support the efficient 
selection of mathematical expressions.  \citet{ref:bier2006entityquickclick} 
designed a technique for rapid selection of entities in text (such as addresses) with a 
single click. The symbol selection mechanism in ScholarPhi can be seen as a 
combination of these two features, supporting single-click selection of 
mathematical expressions, with refinement of the selection to choose specific 
sub-symbols of that expression via additional clicks. In the future, ScholarPhi 
may support the efficient selection of many nonce words at once in a passage using 
fuzzy text selection techniques such as those proposed by 
\citet{ref:hinckley2012informal} and \citet{ref:chang2016supporting}.

\subsubsection{Information Highlighting and Fading}
ScholarPhi is  designed to support the efficiency of visual querying present in contemporary  code editors like VSCode~\cite{ref:vscode}, in which arbitrary text (i.e., a variable  or expression) can be selected, and all other appearances of that same text are 
instantly highlighted everywhere else in the text. In the design of its lists of  definitions and usages, ScholarPhi also draws inspiration from tools such as  LiquidText~\cite{ref:tashman2011liquidtext}, which supports viewing lists of  within-text search results side-by-side with the query term highlighted. In its design of the ``declutter'' filter, ScholarPhi draws on the design of 
visual filters already present in prototype and production tools. The fading out  of content in order to direct a reader's focus to information of interest is a 
design pattern that has been used in interactive 
tutorials~\cite{ref:kelleher2005stencils} in which instructions are highlighted while the rest of the user interface is faded, as well as in interactive debugging  tools~\cite{ref:ko2009finding,ref:dragicevic2011gliimpse}.

\subsubsection{Readability versus Document Augmentations}
On the whole, evidence has supported the usefulness of embedding explanations in texts.  
In the context of second-language learning, embedded glosses for unfamiliar 
vocabulary have been shown to lead to vocabulary 
learning~\cite{ref:yanagisawa2020different}, and improved 
comprehension~\cite{ref:taylor2006effects}.

That said, in making texts interactive, there is a key tension between assisting 
the reader and distracting them. On the one hand, studies such as one run by 
\citet{ref:rott2007effect} suggest that the best comprehension outcomes can be 
achieved when all words that have glosses are marked. On the other hand, interactive 
texts change a reader's behavior. Understandably, readers are more likely to click 
on words that are visibly interactive~\cite{ref:deridder2002visiblechi}, leading 
to what has been called by some ``click-happy 
behavior''~\cite{ref:roby1999whats}. Furthermore, studies of texts augmented 
with hyperlinks have sometimes shown that these augmentations lead to worse 
comprehension of the texts, rather than better 
comprehension~\cite{ref:destefano2007cognitive}. What the evidence suggests 
overall is that amidst the appeal of interactive reading interfaces, great care 
must be taken during design to make sure not to introduce features that will 
ultimately distract readers from the cognitively demanding task of reading.

\subsection{Tools for Reading Scientific Papers}

\subsubsection{Links to External Resources}

Tools can help researchers read scientific papers in a number of ways. 
To reduce the need to click away from the paper currently being read, 
some online journals now allow readers to view metadata by clicking on citations~\cite{ref:elifelens,ref:pubmed,ref:springer}. 
Experimental tools have been built that
augment papers with additional information about cited 
papers~\cite{ref:powley2009enriching}, bias in study design~\cite{ref:marshall2016robotreviewer}, and links to external learning 
resources~\cite{ref:liu2015scientific,ref:jiang2018mathematics}. They have 
supported interpersonal explanations, allowing peer 
reviewers~\cite{ref:peerlibrary}, collaborators~\cite{ref:yoon2014richreview},
instructors~\cite{ref:mccartney2018annotated}, 
strangers~\cite{ref:fermatslibrary}, and
crowds~\cite{ref:jiang2014crowdhilite} to annotate and discuss
passages of papers.
Other approaches to saving the scientist time include tools to  support literature search (e.g.,~\cite{ref:zhang2008citesense,ref:ponsard2016paperquest,ref:qian2019beyond}),   summarize text ~\cite{ref:scholarcy,ref:cachola2020tldr}, or 
rewrite passages in simpler language~\cite{ref:kim2016simplescience}. 

\subsubsection{Links within Papers}
Reading interfaces can also assist researchers by helping them navigate to 
information of interest within a paper. For several years, interfaces for reading PDFs 
have provided standard affordances for jumping within a paper using hyperlinks. Typesetting software 
like \LaTeX{} can automatically embed clickable links from references to figures, 
equations, and sections to the content they refer to, and from citations to 
bibliographies. 

Prototype tools have been built to further assist readers in 
finding passages about topics of interest~\cite{ref:graham1999reader}, in 
jumping between a passage that describes research results to the relevant parts of data tables ~\cite{ref:kong2014extracting,ref:kim2018facilitating,ref:badam2018elastic}, and in jumping to passages that answer  their natural language questions~\cite{ref:zhao2020talk}.  Other tools has augmented static figures in papers with animated~\cite{ref:grossman2015your} or interactive~\cite{ref:masson2020chameleon} overlays.

Of particular relevance to this paper are experimental systems that 
surface explanations of terms and symbols in scientific papers. Tools have been 
developed that link from terms to the pages that define them on 
Wikipedia~\cite{ref:abekawa2016sidenoter}, 
and which link from key phrases in papers to topic pages where those phrases 
are defined alongside excerpts about those topics from other 
papers~\cite{ref:sciencedirect}. 

\subsubsection{Tools for Reading Math}

In response to the unique challenges of reading mathematical texts, prototype 
tools have been designed to expose definitions of math expressions within a
text~\cite{ref:alcock2009eproofs,ref:pagel2014mathematical,ref:kohlhase2018discourse}. e-Proofs provide guided tours of proofs, selectively fading parts of the proof that are not currently the focus of the tour
~\cite{ref:alcock2009eproofs}.  e-Proofs were designed to augment
single-page proofs rather than papers. The Planetary system
lets readers look up the meanings of operator symbols in external knowledge 
bases, and reveals simplified versions of equations with details 
elided~\cite{ref:kohlhase2011planetary}.

Studies of the e-Proofs system (see~\cite{ref:roy2014evaluating,ref:roy2017multimedia})
hint at design tensions in tools for reading math. It was found that while readers used the tools of their own 
accord~\cite{ref:roy2014evaluating}, many features that were introduced to 
assist readers, such as audio walkthroughs of the content, got in readers' 
ways~\cite{ref:roy2017multimedia}.  ScholarPhi consolidates and
extends features from these prior prototypes, and introduces additional features and affordances, with the goal of helping readers  understand mathematical symbols among other nonce words.

%% file: 03-reading-study.tex
\section{Design Motivations}
\label{sec:design-motivation}

\begin{changes}
The design of ScholarPhi is motivated by insights from an iterative design 
process. This section reports insights arising from a formative study, a 
review of the related work, prototyping efforts, and informal 
usability studies of early prototypes.
\end{changes}

\subsection{Formative Study}
\label{sec:formative}

To better understand how the presence of nonce words affects the reading experience, 
we conducted a small formative study. Nine readers (four graduate students, five 
undergraduate students, referred to as R1--9 below) participated in an 
observational study in which they read a scientific text of their own choice.  
Six participants brought research papers (R1--5, R8); five of these papers were 
about computer science and one was about architecture. Three participants 
brought instructional texts on the topics of data science (R6), experimental 
design (R7), and formal analysis (R9).  

Participants were asked to read their text for forty minutes and simultaneously think aloud.  Readers reported when they encountered confusing passages of 
text and  described whether they intended to look up information to 
clarify their confusion. If they chose to look for such clarifying information, 
they described where they looked and why. Our findings were as follows:

All but one reader expressed confusion at a term used in the text (R1--3, 
R5--9). In some cases, the confusion was about a term that was specific to the 
scientific discipline of the text (R3, R5--9), such as the terms ``diacritic'' 
(R3) or ``population parameter'' (R6). For papers from computer science, such 
terms included  benchmarks used to test an algorithm (R3) and 
baselines against which an algorithm of interest was compared (R5).

In other cases, the terms causing confusion came from within the same paper.  
Authors introduced terms to describe their methods (``symbolic validator'' (R1), 
``backtranslation'' (R3)) whose meanings readers could not surmise when viewed 
apart from their definitions. Authors would invent shorthand for running examples 
(e.g., a test set of cow images named ``cow'') that they then referred to 
by that shorthand (i.e., ``cow'') in figures, which could be confusing if the 
reader was reading the text out of order (R5). Texts could also be sprinkled 
with vague back-references to assumptions (R5), analyses (R6), parameters (R8), 
and
theorems (R9) that readers could not recall. In some cases (R6, R8), readers 
were not sure whether a term referred to a passage in the current text or 
in another text.

Mathematical symbols were another source of confusion (R2--4, R6). 
Readers sometimes simply could not understand the meaning of a symbol (e.g., ``$\Theta_s$,'' ``$M$,'' 
``$p$,'' ``$q$,'' ``$x$,'' ``$y$,'' ``$y_1$,'' R2--4, R6).  In other cases, they 
wanted information about how a set of symbols were used in combination. For 
example, R4 scanned the appendix of the research paper they were reading to 
better understand  the meaning of a ratio ``$M / N$'' that appeared in one 
of the equations. Readers also wondered about the values that symbols were 
assigned (R2, R3, R6). For example, one reader (R2) wondered what value  the 
regularization parameter $\lambda$ was set to when a model was trained. Another 
reader (R3) wanted to see example data that could be used as inputs $x$ and $y$ 
to a translation algorithm.

Thus, confusion about terms and symbols (nonce words, in our terminology) was common among the readers in the 
study. Readers' strategies for resolving this confusion varied based on how 
important it was that they understood a nonce word.  If it seemed important, a reader often 
attempted to infer meaning from context (R3, R6--9). If they could not surmise 
the meaning from context, readers would sometimes delay looking up an 
explanation with the hope that they might find one later in the text (R1, R3, 
R4, R6--9). A drawback of this approach, described by R1, is that a reader may 
reach a point in the text where they lack an understanding of so many important 
terms that they can no longer understand the text without stopping and searching for explanations.

\begin{figure}
\centering
\includegraphics[width=.47\textwidth]{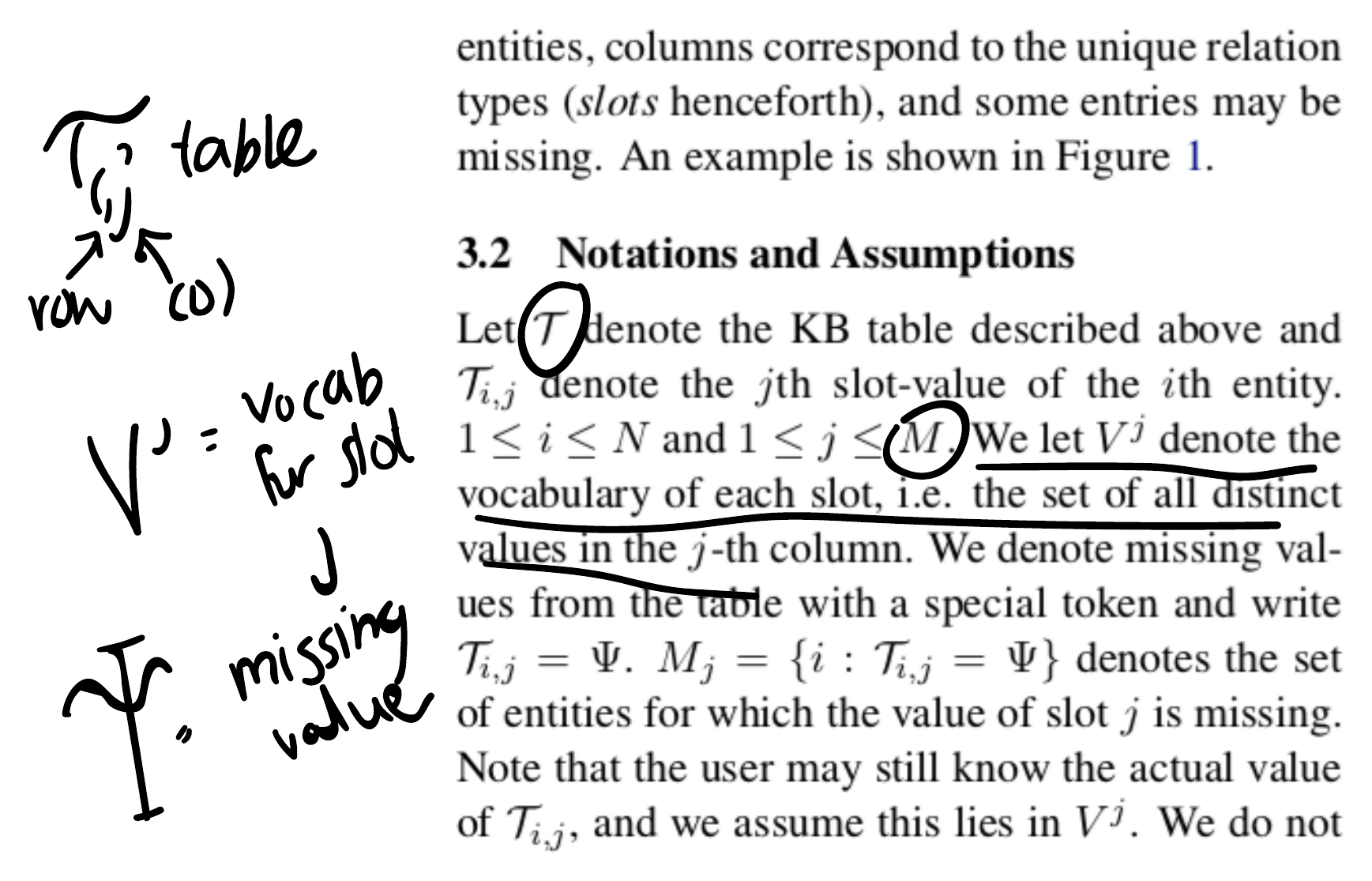}

\caption{When researchers have trouble understanding nonce words, they look up 
explanations elsewhere. \textmd{One researcher in the formative study 
proactively assembled glossaries in the margins of the paper for key symbols 
(above). The researcher annotated the text with definitions of symbols and miniature equation 
diagrams (see the annotation for $\mathcal{T}_{i,j}$).}}
\label{fig:personal_glossary}
\Description{%
An excerpt from a scientific paper with the handwritten annotations of a 
participant in the formative study.

Of note is that the participant drew three symbols in the margin and wrote brief 
definitions next to each one. The symbol ``Tau-sub-i-comma-j'' is defined as 
``table.'' The participant drew arrows to the ``i'' and ``j'' subscripts, and 
wrote definitions for each subscript (i.e., ``row'' and ``col''). The 
participant also wrote definitions for two other symbols in the same margin.
}

\end{figure}

Eventually, many readers needed to do just that, and stopped reading in order to look up 
explanations. One participant referred to this as an undesirable ``context 
switch'' which takes them out of the ``headspace'' of understanding a 
complicated passage (R4). When looking for explanations, five readers looked 
elsewhere in the same text (R2-4, R8, R9). This entailed backtracking within the 
text (R3, R4), jumping forward (R2, R4), opening within-text glossaries (R8), 
and performing within-text search (i.e., ``Control-F'' search) within the 
reading application (R9). Those reading instructional texts often consulted 
external references like web search results (R6, R8), dictionary applications 
(R7), and Wikipedia (R9). One reader took a proactive approach to reducing the 
cost of within-paper lookups by assembling glossaries for key symbols in the 
margins of the text (R4, see Figure~\ref{fig:personal_glossary}). 

This study indicated that readers of scientific papers, and scientific texts 
more generally, frequently have questions about nonce words. To answer these 
questions, readers either infer answers from context, wait for an answer, or 
look for explanations elsewhere. While readers do look for explanations 
elsewhere, they try to avoid doing so as it takes them away from the text they 
are trying to understand. These observations suggest that readers could benefit 
from interfaces that make explanations of nonce words, and unfamiliar terms more generally, available to them without 
distracting them from the task of a careful reading.

%% file: 04-design-motivations.tex
\begin{changes}
\subsection{Design Process}

The design of ScholarPhi was refined through an iterative design process lasting 
twelve months. Our research followed a process similar to research through 
design~\cite{ref:zimmerman2007research}, consisting of iterative ideation, 
prototyping, and assessment. This process yielded a multifaceted design space, 
representing choices that must be made when designing an interactive tool that 
shows definitions of nonce words.

Design alternatives were identified by reviewing literature reviews on 
e-glossaries (e.g.,~\cite{ref:roby1999whats,ref:yanagisawa2020different}) and
research and commercial tools (see Section~\ref{sec:related-work}), and
by brainstorming within our team and with users of early prototypes. The design 
space appears in Table~\ref{tab:design_space_figure}. 

\input{tables/design-space-figure}

Five prototypes were developed and assessed. The last of these prototypes is the 
tool we call \emph{ScholarPhi}, and is described in Section~\ref{sec:demo} and 
assessed in Section~\ref{sec:evaluation}.
The first four prototypes were designed to evaluate promising design 
alternatives. Table~\ref{tab:design_space_figure} indicates which features were 
present in early prototypes, and in the final design.
\end{changes}
Each prototype was evaluated in a pilot study of its own:

\begin{itemize}
    \item Study $D$ (Declutter lens only): 4 researchers (D1--4)
    \item Study $S$ (Side notes containing  definitions, defining formulae, and usages): 4 researchers (S1--4)
    \item Study $T$ (Tooltips instead of side notes): 9 researchers (T1--9)
    \item Study $E$ (Equation diagrams and a complex version of tooltip interaction flow): 9 researchers (E1--9).
\end{itemize}

The first two studies ($D$ and $S$) were observational studies. Participants 
thought aloud as they used the tool. For the second two studies ($T$ and $E$), 
participants read on their own, participated in a 30-minute focus group 
discussion, and filled out a questionnaire about their experience. Seven
participants in these last two studies (T1--3, E1--4) participated in a 
15-minute follow-up interview. In each study, participants read a different 
scientific paper. \change{Two researchers (S2, S3) participated in multiple 
studies}.

One author analyzed transcripts from all studies following a qualitative 
approach. This yielded the following \change{seven} design motivations for 
designing effective interfaces for providing in-situ definitions within 
scientific texts.

\subsection{Design Motivations}
\label{sec:design-motivations}

\paragraph{M1. Tailor definitions to the location of appearance}

The same nonce word can have multiple conflicting definitions throughout a paper.   
For example, in the paper used as stimulus in the formal 
study~\cite{ref:strubell2018linguistically}, the symbol $T$ took on multiple 
distinct senses including  referring to the dimensionality of a vector $x_t$, 
being part of a composite symbol $T^{(j)}$ used to refer to a layer in a neural 
network, and being used as the matrix transposition operator in several display 
equations. When readers used a prototype that showed definitions of \emph{all} 
of these senses in a list, they wanted to know which ones were the most 
appropriate to  the passage that they were reading (S1--3). Readers requested 
that the tool   show the definitions appropriate to the place where they asked 
for them (S1).  They also asked to see the context surrounding a definition (S2, 
S3).

A related principle is eliminating redundant definitions. If a reader selected a 
nonce word within a passage where it was being defined, they did not wish to see a 
tooltip containing the definition sentence they were already reading (S1, T9).

\paragraph{M2. Connect readers to definitions in context}

Four readers requested the ability to jump from a definition to the passage 
where it appeared in the paper (S1--3, T5, T6).  This would help them judge the 
relevance of the definition (S1--3) and assess what they suspected may be incorrectly
extracted definitions (T5).

\paragraph{M3. Consolidate information}

While the information that explains a nonce word can be scattered across a paper, 
readers want explanations that consolidate all of that information in one 
compact, concise passage. When they clicked on a composite symbol, they wanted 
to see explanations of each sub-symbol that made up the symbol (E2, E4). They also 
expected the interface to be able to gather explanations for semantically 
similar symbols that differed in their surface features, such as
showing a definition for ``$\textrm{PMA}(\cdot{})$'' that was extracted for the 
function ``$\textrm{PMA}(X)$'' (E1). 

\paragraph{M4. Provide scent}

In all prototypes, nonce words were marked with a light dotted underline. Readers
appreciated that the underlines provided scent of which words they could click 
to see definitions (S2--4). Participants did not turn off this feature, although 
they were provided with this option in later versions of the design.

\paragraph{M5. Minimize occlusion}

In two prototypes, tooltips were packed with definitions, defining formulae, and 
usages for symbols. Readers reported that these tooltips occluded text that they 
wished to see (T4, T6, E7) without providing much value beyond the first 
definition (T1, T4--6). Still, some readers desired tooltips as opposed to side 
notes, as it allowed them to view definitions without losing their place in the 
text (E3, E4). The current prototype attempts to balance these conflicting needs 
by providing a compact tooltip that contains only the most recent definition of 
a nonce word and a few small buttons for accessing lists of definitions, defining 
formulae, and usages. A tooltip for a nonce word can be hidden by clicking on a 
``close'' button within the tooltip.

\paragraph{M6. Minimize distractions}

The user interface was revised several times to remove features that, while originally
envisioned as being helpful, distracted from the reading task. One reader aptly 
described, ``I was trying to pay more attention to the paper than the tool and 
the paper requires a lot of overhead to understand.  So I didn't have much left 
over for the tool'' (E1). One prototype used several highlighting colors to 
indicate appearances, usages, and definitions of a selected nonce word; however, this 
added visual clutter that was hard to understand (E3). The current prototype 
uses a single static highlight color. Readers were asked across multiple studies 
whether they found underlines beneath the nonce words distracting. They repeatedly 
reported that they did not  (S2--4, T5, T7).  However, one reader did request 
the ability to turn them off (E1), which has been included in all recent 
prototypes of the interface.

\begin{changes}
\paragraph{M7. Support error recovery}

The systems that are used to detect definitions in scientific papers are prone  to error; user interfaces that build off of this technology must take this into account. 
Drawing on guidelines from the literature on 
interacting with intelligent interfaces when there are errors (synthesized by Wright et al.~\cite{ref:wright2020comparative}), the ScholarPhi interface makes use of the following guidelines:
 provide paths forward, and support
efficient dismissal. 

Other guidelines that appear in the literature, such as promoting user 
corrections of errors and displaying auto-detected errors, may impose 
distractions while reading, and so were not incorporated into the current 
version of the interface, although future work may determine that these 
approaches are helpful.

\end{changes}

%% file: tables/design-space-figure.tex
\begin{table}

\centering
\includegraphics[width=0.47\textwidth]{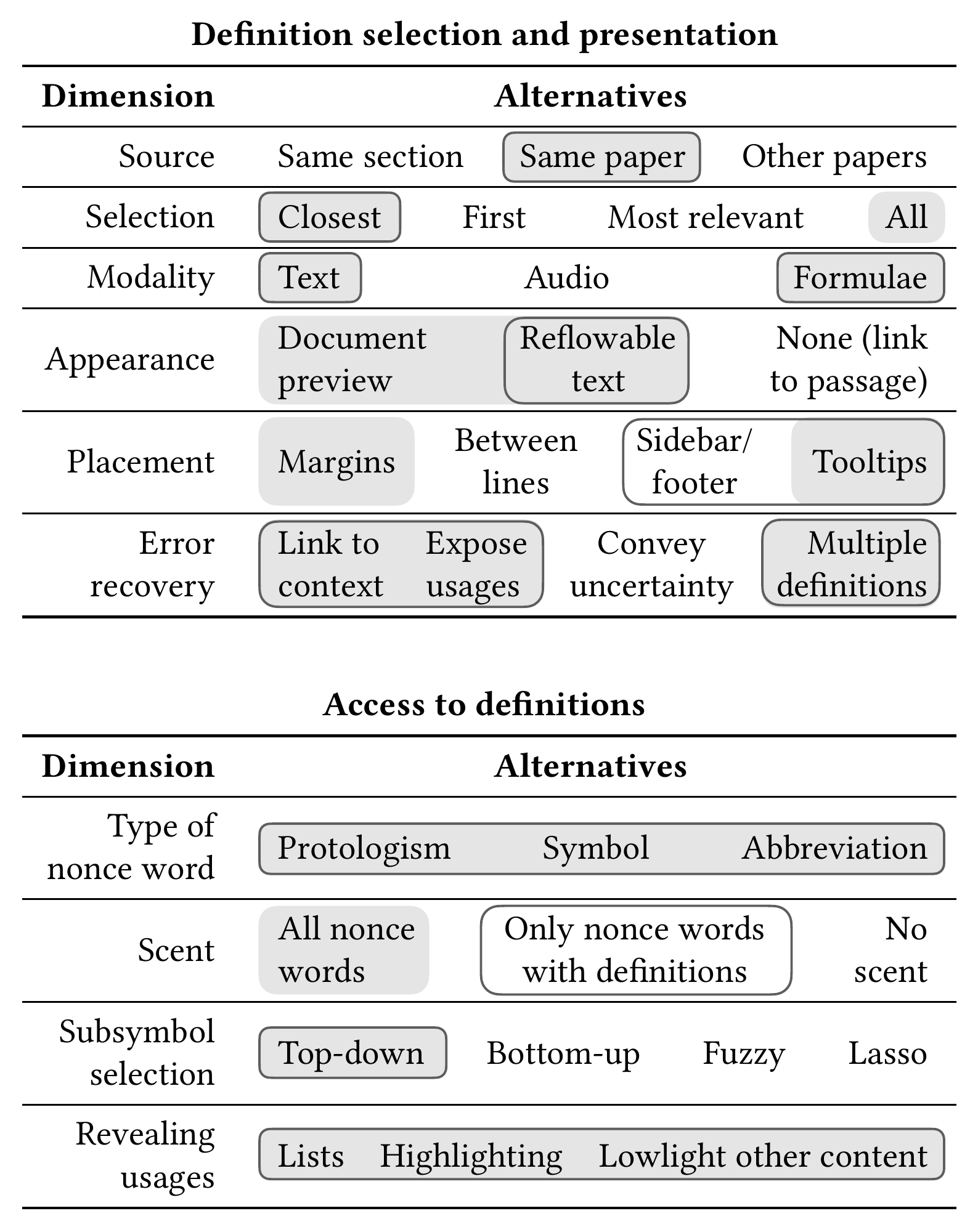}
\caption{\change{Design alternatives for applications that reveal definitions of 
terms and symbols. \textmd{Alternatives tested in our early prototypes are 
highlighted in gray. Those selected for inclusion in our final design are 
circled with a solid border.}}}
\label{tab:design_space_figure}
\Description{%
Table of design alternatives. Highlights and borders indicate which alternatives 
were tested in early prototypes, and which were implemented in the final design.

The following alternatives were assessed in early prototypes:

For definition selection and presentation:

Source: Same paper

Selection: Closest, All

Modality: Text, Formulae

Appearance: Document preview, Reflowable text

Placement: Margins, Tooltips

Error recovery: Link to context, Expose usages, Multiple definitions

For access to definitions:

Type of nonce word: Protologism, Symbol, Abbreviation

Scent: All nonce words

Subsymbol selection: Top-down

Revealing usages: Lists, Highlighting, Lowlight other content

The following dimensions were chosen for the final design:

For definition selection and presentation:

Source: Same paper

Selection: Closest

Modality: Text, Formulae

Appearance: Reflowable text

Placement: Sidebar / footer, Tooltips

Error recovery: Link to context, Expose usages, Multiple definitions

For access to definitions:

Type of nonce word: Protologism, Symbol, Abbreviation

Scent: Only nonce words with definitions

Subsymbol selection: Top-down

Revealing usages: Lists, Highlighting, Lowlight other content
}
\end{table}

%% file: 05-system.tex
\section{User Interface}
\label{sec:demo}

We illustrate the experience using ScholarPhi through a set of four
scenarios, where a reader wishes to know the meaning  of a specific nonce word.  
Each scenario is chosen so that one of ScholarPhi's features is uniquely 
well-suited to the reader's task. 

To explain the design decisions underlying a feature, we refer back to findings 
from the formative research. Specifically, we note whenever a design choice was 
informed by one of the design motivations M1--7 that were introduced in 
Section~\ref{sec:design-motivations}.\footnote{
\change{The papers in these scenarios are recent computer science papers by Lee et al.~\cite{ref:lee2019deep} and Strubell 
et al.~\cite{ref:strubell2018linguistically}. The latter paper, of which large passages are shown,
is published in the EMNLP 2018 proceedings under 
a Creative Commons ShareAlike-4.0 License.}
}

\subsection{Definition Tooltips}
\label{sec:definition-tooltips}

When a reader wants to know the meaning of a nonce word, ScholarPhi lets them look up 
the meaning by clicking the nonce word. This reveals a definition tooltip (see  Figure~\ref{fig:teaser}).

Definition tooltips appear directly beneath the selected nonce word. This 
placement is intentional. By placing the definition beneath the word, as opposed 
to placing it in a document margin or a glossary elsewhere in the text, a reader 
need not divert their gaze from the text. In this way, the tooltip placement is 
chosen to minimize distraction (M6). Furthermore, to avoid occluding the text 
(M5), tooltips are compact. Their dimensions never exceed half the page width, 
nor are they permitted to be longer than four lines tall.

If there are multiple definitions of a nonce word available within the paper, 
ScholarPhi shows the definition that it infers as being most relevant to the 
context. Specifically, it uses a heuristic of showing  the definition that 
appeared most recently before that appearance of the word.  This reduces mental 
effort that seeing multiple definitions over the nonce word would incur (M1) and 
reduces the amount of text occluded by the tooltip (M5).

\change{For instance, in the following passage from Lee et al.~\cite{ref:lee2019deep}, $k$
initially refers to an index of a component in a mixture of Gaussians.}

\inlinefigure{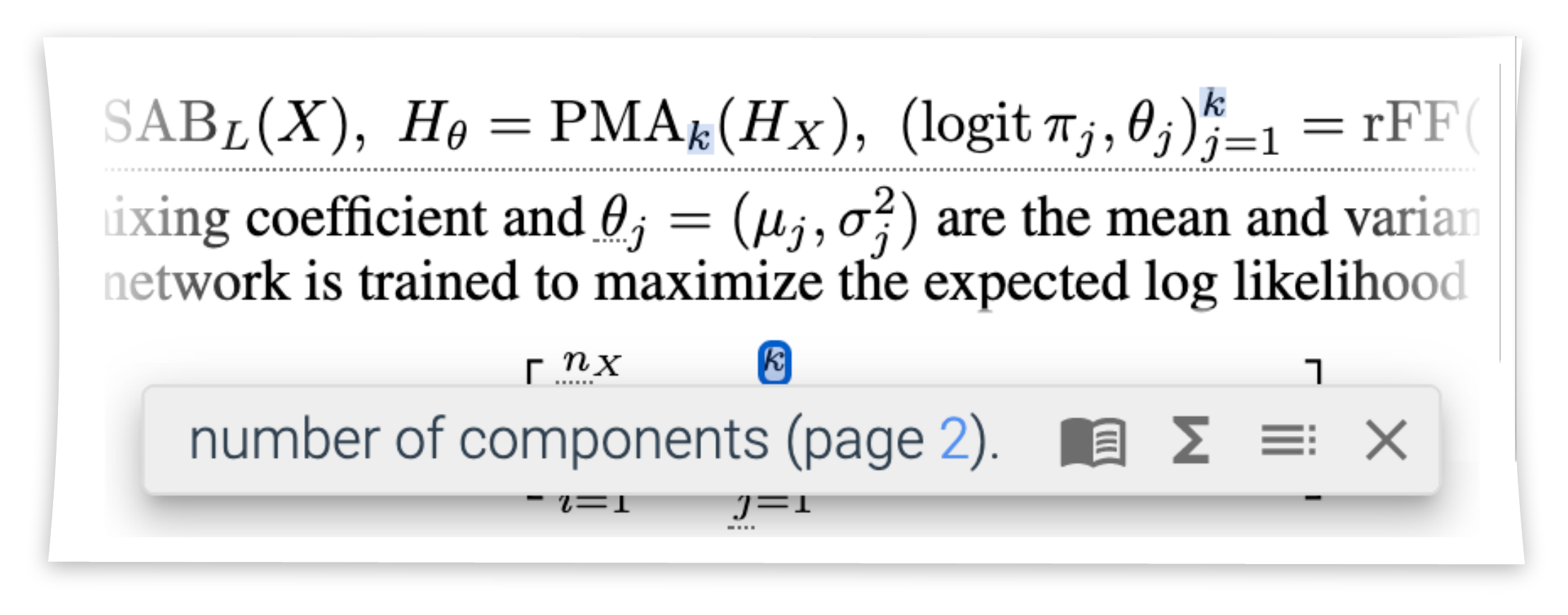}

However, in a later passage, $k$ is given an entirely different meaning---a 
parameter that controls the number of clusters output by a clustering algorithm.  
When the reader opens a tooltip in this other passage, they again see the 
appropriate definition.

\inlinefigure{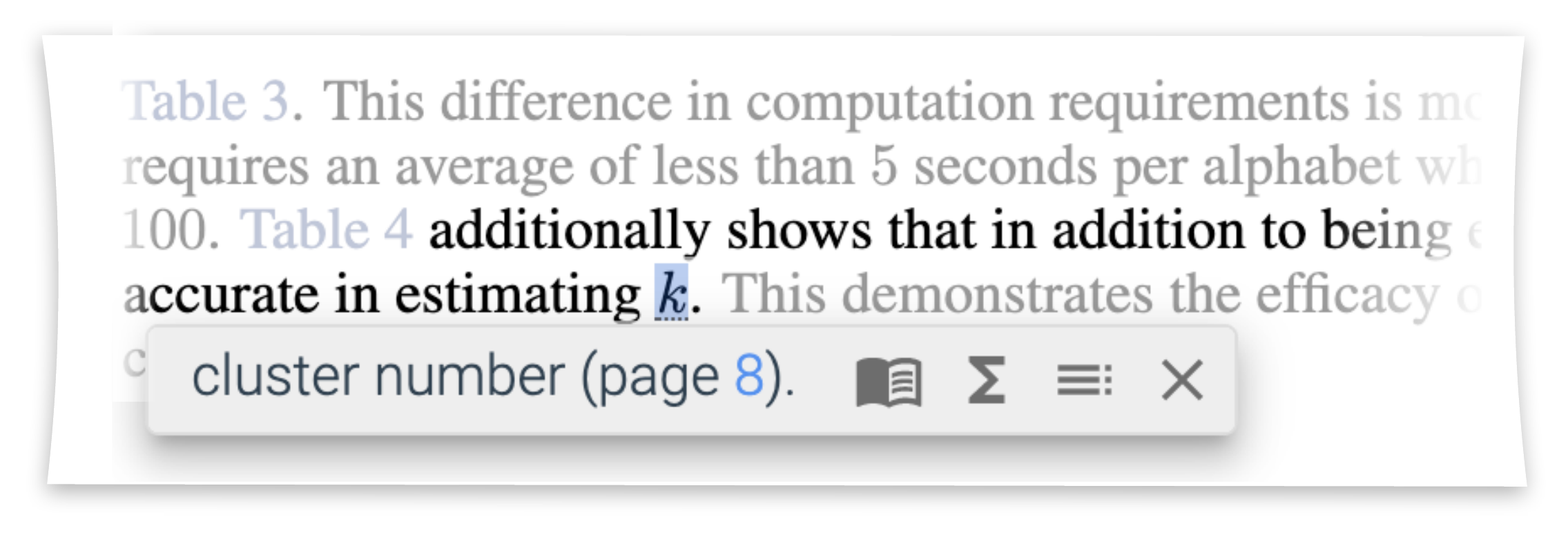}

After seeing a definition in the tooltip, a reader may want more information 
about the nonce word. For instance, they may want to know whether the authors 
recommended that a specific number of $k$ components be used in the mixture of 
Gaussians. To help the reader answer questions like this, ScholarPhi connects 
the reader to definitions in context (M2). The reader can view the definition in 
context by clicking the hyperlink next to the definition (e.g., ``page 2'').
ScholarPhi scrolls the paper to the definition, highlighting 
the sentence that the definition came from:

\inlinefigure{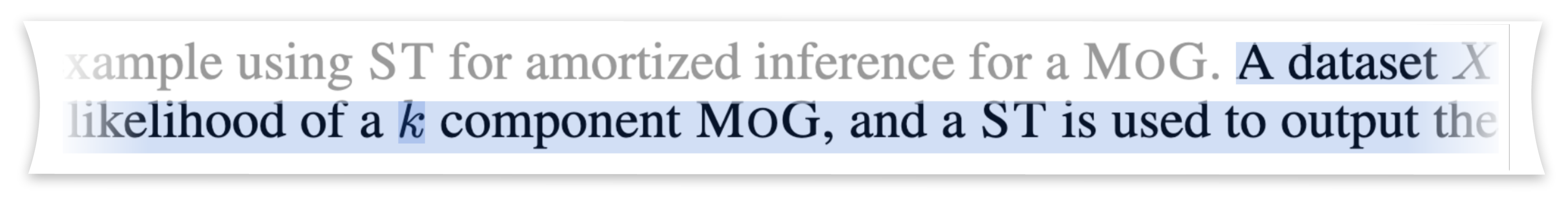}

When the reader has finished consulting the highlighted passage, they can click 
their web browser's ``Back'' button to return to the definition tooltip at their 
previous position in the document.

\paragraph{Lists of usages}
A reader can also look for more information about a nonce word by reviewing the 
usages of the word. To connect a reader with these usages, the definition 
tooltip provides three buttons. The buttons let a reader open lists of all prose 
definitions of the word, all defining formulae (i.e., formulae in which the 
nonce word appears on the left-hand side of an assignment), and all usages 
(i.e., passages that refer to the nonce word). Together, the buttons provide a 
way for readers to access a consolidated collection of everything that 
ScholarPhi knows about a nonce word (M3).

When a reader clicks a button, the corresponding list opens in 
a dedicated sidebar, rather than in the tooltip, to avoid occluding the text 
(M5).
Each usage in the list comprises one sentence referring to the nonce word and a 
link to the sentence where it appears in the paper (M2). To help readers 
evaluate the relevance of a usage among 
dense text and equations, the nonce word is highlighted.

\inlinefigure[.46\textwidth]{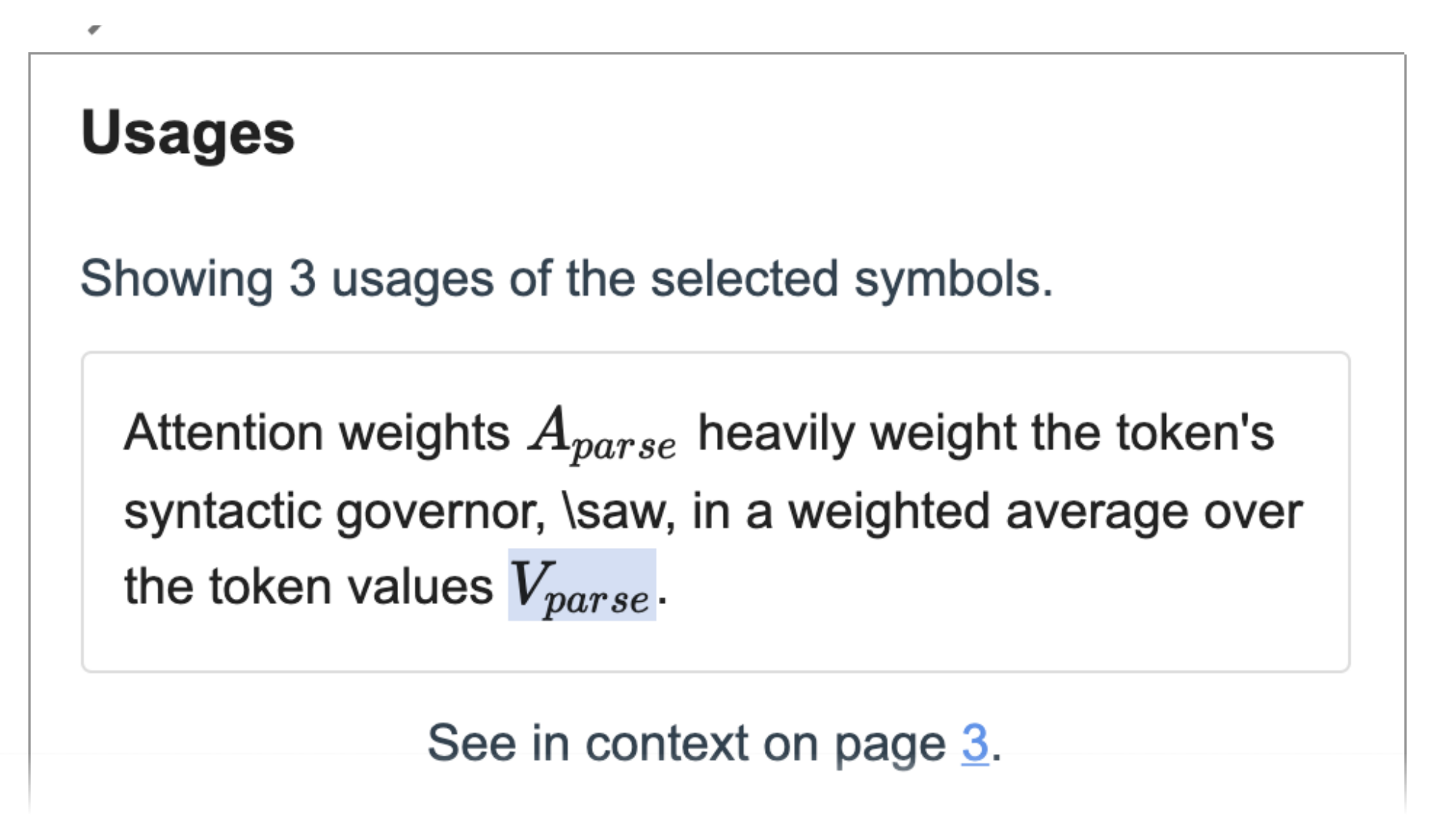}

The tooltip buttons for opening usage lists provide scent to help 
a reader understand how a nonce word is defined and used (M4). By hovering over 
a button, the reader can see how many definitions, defining formulae, or usages there are for a nonce word. A
button is disabled if no definitions, defining formulae, or usages exist.

To avoid disorienting the reader, a tooltip always makes the same information 
available to a reader in the same layout: buttons for lists of definitions, 
defining formulae, and usages, as well as a definition if one is available. If a 
tooltip is opened for a nonce word within the sentence where the word is 
defined, the definition tooltip reports, ``Defined here.'' This way, tooltips do 
not distract the reader from the text with a definition they have already seen, 
or are about to see (M6). If no definition exists for the nonce word, then the 
three buttons to access the usage lists are still shown, but those with no 
information behind them are grayed out.

\paragraph{Scent}
While some nonce words are defined in a paper, others are not. Authors may assume the 
meaning of a nonce word is implicit or they may simply forget to define it.  ScholarPhi provides visual scent~\cite{ref:pirolli1999information} 
to help readers determine whether they'll find a definition for a nonce word 
before they click on it (M4).  This visual scent is provided in the form of a 
subtle dotted gray underline beneath the nonce word.  For instance, 
in the following passage, readers can open definition tooltips for any of the 
underlined nonce words, ``CoNLL-2005,'' ``SRL,'' or ``LISA.''

\inlinefigure[.4\textwidth]{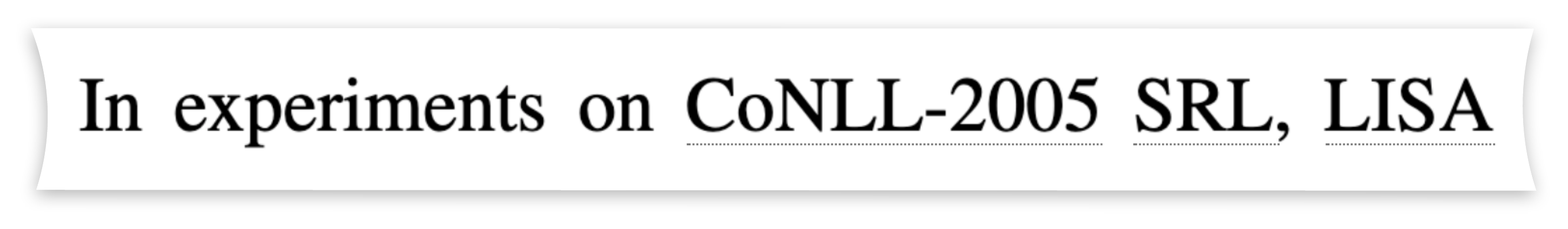}

So that it does not divert a user's attention from the text needlessly (M6), 
ScholarPhi assumes that a reader will not want to view a nonce word in a sentence 
that defines it, and so does not underline these occurrences.
The rules for underlining symbols are more nuanced. Papers can 
contain composite symbols where certain sub-symbols (e.g., subscripts or 
superscripts) are defined, but the composite symbol as a whole is not. In such a case,
ScholarPhi highlights sub-symbols for which definitions are available. In the 
passage below, ScholarPhi highlights symbols to indicate that definitions are available for 
``$t$,'' ``$\mathcal{X}$,'' and ``$r_t$.'' Because the composite symbol 
``$y_t^{prp}$'' is defined in the sentence, it is not underlined.

\inlinefigure[.45\textwidth]{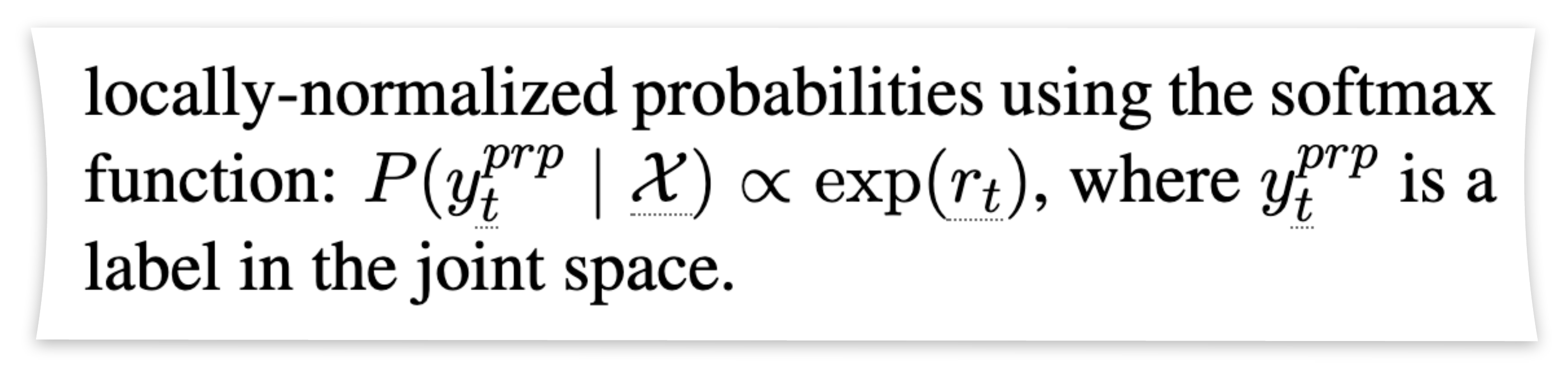}

\paragraph{Symbol selection}
In a conventional interface for reading papers, one challenge to searching for 
information about a symbol is simply selecting the symbol. Because the text for 
a symbol is often split across multiple baselines (i.e., in subscripts or 
superscripts), conventional text selection mechanisms may fail to select 
precisely those characters that belong to the symbol. To reduce the cost of 
accessing explanations, ScholarPhi supports efficient selection of symbols.
Symbols can be selected by clicking them once (steps ``1'' and ``2'' below).  
Once a symbol is selected, all sub-symbols that belong to it are highlighted and 
can be selected with a click (``3'').

\inlinefigure[.4\textwidth]{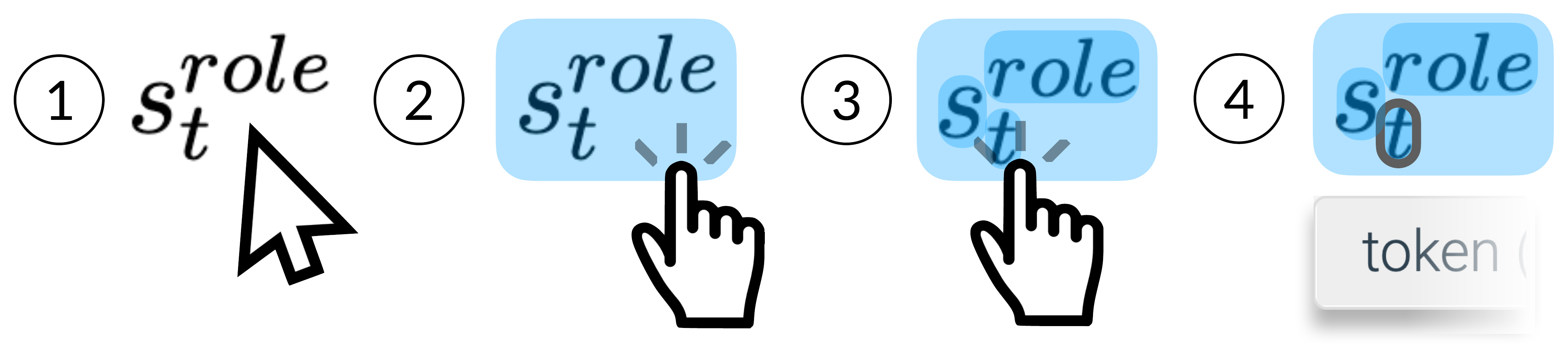}

By helping readers rapidly select sub-symbols, it is hoped that ScholarPhi lets
readers understand the meaning of a composite symbol in terms of the meanings of 
its parts (M3).

\subsection{Declutter}
\label{sec:declutter}

To help readers quickly
find information about a nonce word that is scattered across a paper, ScholarPhi 
provides a novel feature called ``decluttering.'' When a reader selects a nonce word, 
ScholarPhi ``declutters'' the paper---by highlighting segments of text that 
contain matches, and fading out all other sentences---in an effort to help readers scan the paper for usages.

\inlinefigure[.47\textwidth]{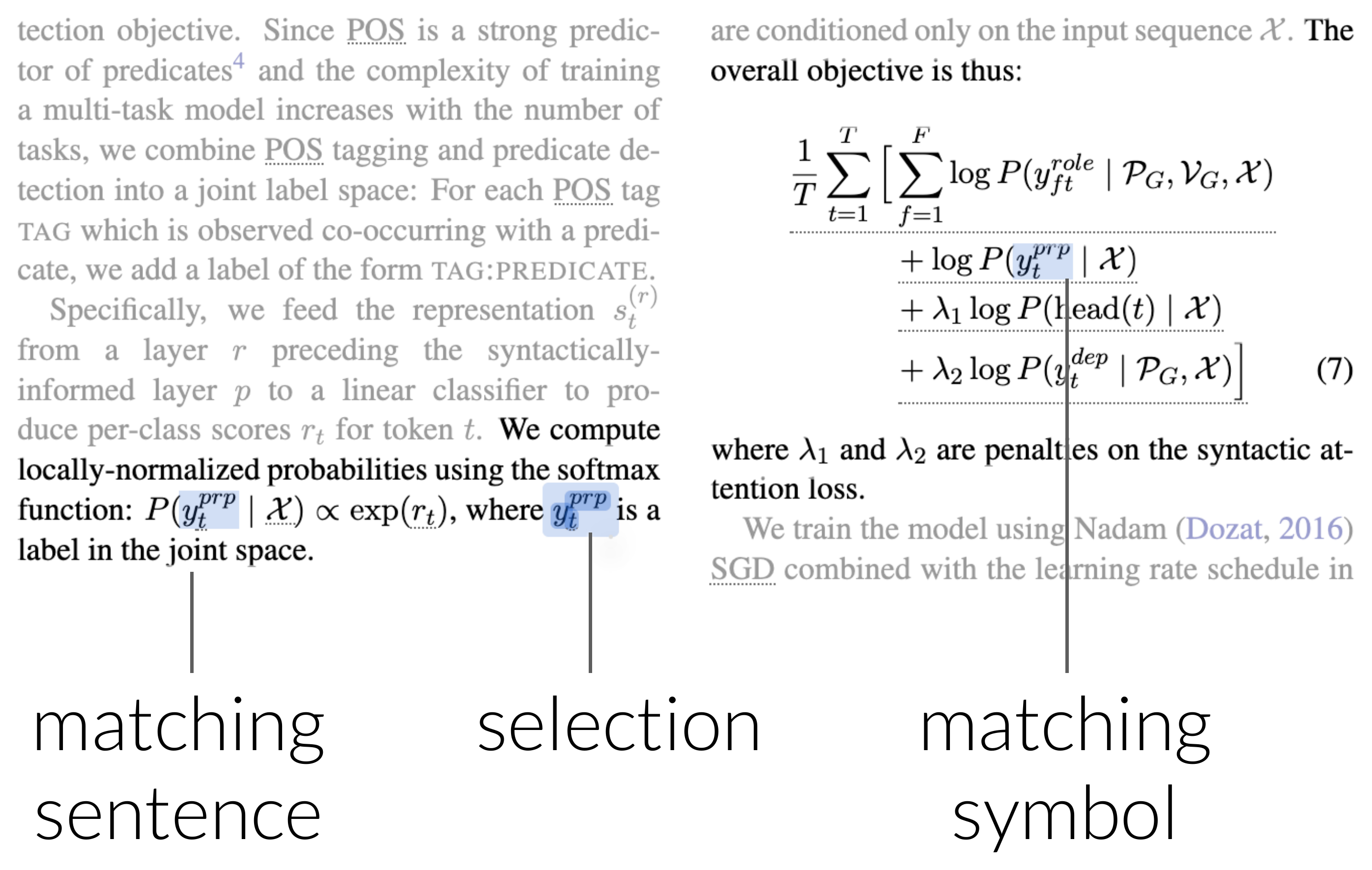}

ScholarPhi provides visual scent (M4) of where usages can be found via a 
conventional search bar. The search bar counts how many times the nonce word appears 
in the paper, and shows the page number of the usage the reader selected. While readers are 
expected to navigate a decluttered document by scrolling through it, the search 
bar also supports navigation between usages with ``Next'' and ``Previous'' 
buttons and arrow key keyboard shortcuts.

Decluttering offers two advantages over the list of usages: it connects readers 
to definitions in context with a view that is grounded in the text (M2) and it 
reduces distractions by hiding irrelevant content, rather than showing 
additional interface widgets (M6). Like the list of usages,
decluttering does not occlude text (M5).

\subsection{Equation Diagrams}
\label{sec:equation-diagrams}

Some passages are rife with nonce words. For instance, tables of empirical results 
are indexed by abbreviations that represent experimental conditions and 
measurements. Equations contain dozens of symbols. For dense passages like 
these, readers may desire the ability to consult the definitions for many nonce 
words at the same time. For display equations in particular (i.e., equations 
that are shown on their own line separated from the text), ScholarPhi provides 
the ability to view definitions of all symbols at the same time. To see the 
definitions of all symbols in a display equation, a reader can click that 
equation. Definitions are affixed to all symbols simultaneously.

\inlinefigure[.47\textwidth]{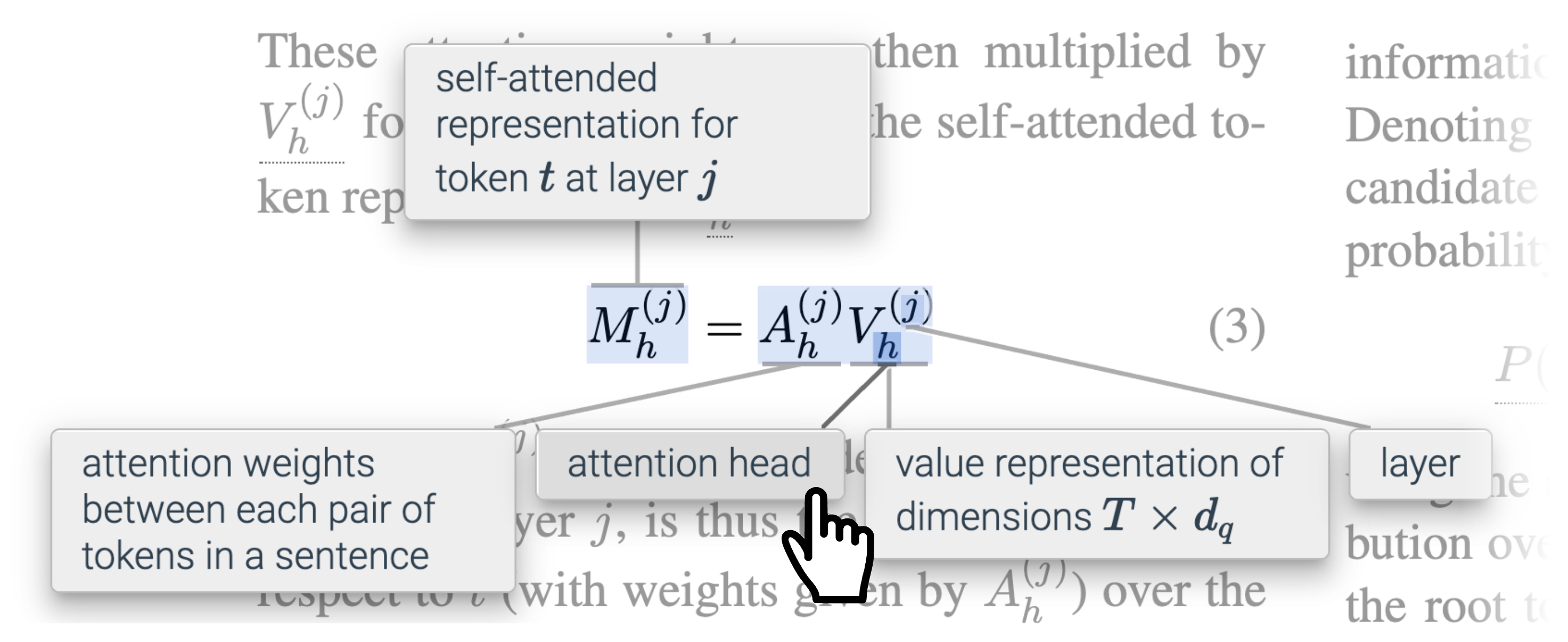}

Definitions are shown for symbols (e.g., ``$V_h^{(j)}$'' in the figure above) and the sub-symbols 
they are composed of (``$h$,'' ``$j$''). Thus, definition information that would 
otherwise be split across multiple tooltips is consolidated into one place (M3).  
Like the definitions that appear in tooltips, the definitions for equation 
diagrams are position-sensitive (M1). By clicking a label for a symbol, a reader 
can open the definition tooltip for the symbol, providing access through the 
definition tooltip to the context of the definition (M2).

Brushing and linking connects the definition labels to the symbols; as a reader hovers 
over a label, the symbol it defines is highlighted with a more saturated 
color than the other symbols. Leader lines connect the definitions to the 
symbols. The leader lines connecting definitions to symbols are diagonal, 
proceeding straight from the definition label to the symbol. This style of 
leader line was chosen as opposed to orthogonal leaders (i.e., leaders 
comprising one horizontal and one vertical segment). While in general, 
orthogonal leaders have been observed to be particularly legible ~\cite{ref:barth2019readability},
we have found that diagonal lines 
stand out better amidst the clutter of other marks in an equation (M6).

\subsection{Priming Glossary}
\label{sec:priming-glossary}

Scientific texts like textbooks often contain glossaries that allow readers to 
look up definitions of terms in a predictable place. One type of glossary that can be particularly 
helpful to readers is what \citet[page 82]{ref:widdowson1978teaching} called a 
``priming glossary,'' or a glossary that is shown to readers before a text to 
help prepare them for problematic words that may appear in the text. ScholarPhi 
prepends a priming glossary to scientific papers. The glossary includes a list 
of key terms and symbols, ordered by their appearance in the paper.

The glossary is intended to help readers in two ways. First, it lets them 
familiarize themselves with the nonce words that will be used in the paper.  And 
second, it provides a reference that can be printed and viewed side-by-side with 
the paper. One advantage to presenting definitions in a priming glossary as 
opposed to tooltips is that definitions for all nonce words can be consolidated 
into one place (M3), allowing a reader to learn about groups of related nonce 
words all at once.  Furthermore, the gloss provides scent (M4) indicating the 
density of nonce words, and the presence of definitions of those words, before 
the reader starts reading.

%% file: 06-implementation.tex
\section{Implementation}
\label{sec:current-implementation}

\begin{changes}

For a given PDF document, the ScholarPhi interface requires information about 
the positions and definitions of the terms, symbols, and sub-symbols within it.
This section briefly describes reference algorithms for obtaining this 
information.\footnote{More details can be found in a forthcoming 
paper~\cite{ref:headdetection}.} The algorithms analyze \TeX{}/\LaTeX{}-based 
PDFs to find exact locations of equations, symbols, and sentences 
(Section~\ref{sec:pipeline}), build up representations of composite symbols 
(Section~\ref{sec:symbols}), and detect definitions for symbols and terms 
(Section~\ref{sec:definitions}). This section also describes the implementation 
of the web-browser-based user interface (Section~\ref{sec:uiimplementation}).

Because most scientific research today is published in PDFs (Portable Document 
Files), the ScholarPhi implementation tackles the challenging problem of 
providing interactions on  PDFs.  It would have been easier to demonstrate the 
technology on HTML or XML format, but that would not have achieved our goal of 
widespread use. For the same reason, we determined that it was important to 
provide the user interface for the document reader directly within a web browser 
without requiring a separate tool to be downloaded.

Algorithms for processing papers are implemented in 10.2k lines of Python code 
and 200 lines of custom \TeX{} coloring macros. The user interface is 
implemented in 10.5k lines of React code. 

\subsection{Domain of Input Documents}

The algorithms below assume that a PDF has been compiled from a manuscript
written in the \TeX{} or \LaTeX{} typesetting language (collectively referred to 
as ``TeX'' below). It also assumes the sources for the manuscript are 
publicly available. This assumption holds for a broad collection of papers in 
computer science, where sources for papers are increasingly hosted on preprint 
servers like arXiv. In fact, arXiv hosts sources for over 1M  
papers~\cite[Section 2.2]{ref:lo2020s2orc}.

The algorithms operate on TeX rather than compiled PDF representations
to improve the precision of detection of inline equations, the segmentation of 
equations into symbols, and the determination of which symbols are children of 
others. With TeX, these tasks become text parsing problems with existing, 
reliable solutions. The dependence on TeX is a stopgap; we anticipate that 
future implementations will accomplish these tasks by processing PDFs 
(e.g.,~\cite{ref:lin2011mathematical}) or images 
(e.g.,~\cite{ref:phong2020hybrid}) of papers directly.

\subsection{Nonce Word Position Detection}
\label{sec:pipeline}

\paragraph{Finding bounding boxes}

To support definition tooltips, equation diagrams, and subsymbol selection, 
ScholarPhi requires bounding boxes for terms, equations, and symbols. To find 
these bounding boxes, our basic approach is to:

\begin{enumerate}

\item Modify the TeX for a paper to assign a unique color to each term, 
equation, or symbol;
\item Compile the modified TeX into a PDF;
\item Render the PDF into images;
\item Analyze the images to detect the colors assigned to each term, equation, 
or symbol.

\end{enumerate}

\noindent
This approach has proven successful in prior work for assembling datasets of 
bounding boxes for figures~\cite{ref:siegel2018extracting} and 
equations~\cite{ref:li2020docbank}.

\begin{figure}
\centering
\includegraphics[width=.47\textwidth]{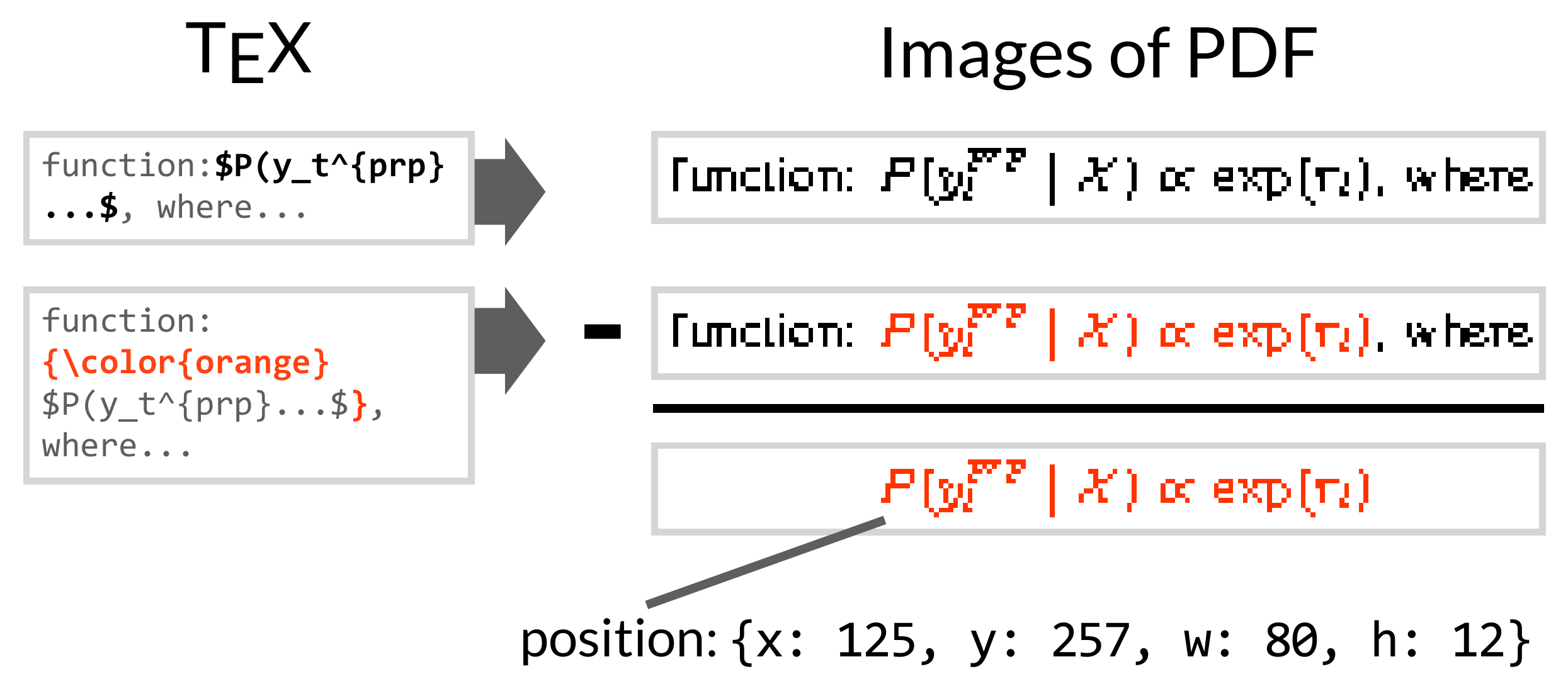}
\caption{Image processing to find bounding boxes of equations, terms and 
symbols. \textmd{Shown is a simple example of our approach. An equation is detected in a paper's TeX  using text
matching rules. The equation is colored using a TeX command. Then, the 
position of the equation is found by differencing the original PDF and the 
colorized PDF and detecting the colored pixels.}}
\label{fig:detection_example}
\Description{%
A decorative diagram showing the process of locating the position of a single 
equation. A description of the process appears in the caption and text.
}
\end{figure}

For example, consider the process of finding the bounding box for a single 
equation (Figure~\ref{fig:detection_example}). First, equations are detected in 
the TeX sources for a paper with regular expressions that match equation 
environments (e.g., pairs of ``\texttt{\$}'' characters). Each equation is 
assigned a color by wrapping it with a TeX color command like 
``\texttt{\{\textbackslash{}color\{orange\}...\}}''. When the TeX is compiled, 
each equation appears in its assigned color.

The position of each equation is found by differencing the colorized PDF with
the original PDF and finding a set of minimal bounding boxes that contain the
assigned color. Boxes are found with a custom blob detector that eagerly creates boxes
surrounding pixels of the same color in the same row of pixels, and then joins boxes appearing
in adjacent rows.
This blob detector
detects the symbol ``$y$'' as one bounding box, ``$i$'' as two bounding boxes 
(with one box for the stem and one for the dot of the ``$i$''),
and a multi-line equation as at least one box per line.

The strength of this approach  over purely image-based or 
PDF-based recognition techniques lies in its ability to find bounding boxes of 
composite symbols, as described below. One limitation of this approach is
that it requires TeX sources to be compiled many times for dense papers. Assuming a paper
contains $N$ nonce words and an image processing library can distinguish
between $C$ different colors, the TeX sources must be compiled at least
$N$ / $C$ times to detect the positions of all nonce words.

\subsubsection{Detecting Composite Symbols}
\label{sec:symbols}

\newcommand{\xml}[2]{\texttt{\textcolor{black}{\textmd{<\textbf{\textcolor{darkgray}{#1}}>\textcolor{black}{\textbf{#2}}</\textbf{\textcolor{darkgray}{#1}}>}}}}

The problem of segmenting TeX equations into symbols is already partially solved  
in open source tools like KaTeX~\cite{ref:katex} and MathJax~\cite{ref:mathjax}, 
which convert TeX equations into structured MathML~\cite{ref:mathml} documents. 
In these documents, nodes often correspond to symbols. For example, the TeX 
equation ``\textbf{\texttt{x\_i}}'' would be parsed into the MathML document: 

\begin{quote}
\centering
\xml{msub}{ \xml{mi}{x} \xml{mi}{i} }
\end{quote}

In this document, \xml{mi}{x} and \xml{mi}{i} represent simple symbols, marked 
as \texttt{mi} or ``identifier'' elements. The document as a whole is one 
composite symbol, consisting of a subscripted symbol with \xml{mi}{x} as the 
base and \xml{mi}{i} as the script.

The KaTeX parser was extended to segment equations. The parser was instrumented 
to produce MathML documents where nodes are annotated with  the positions of the 
characters they represent in the TeX. Then, the MathML document is searched for 
simple and composite symbols. Simple symbols are detected as identifier nodes, 
or rows of identifier nodes that can be merged into one word. Composite symbols 
of three
types are detected:

\begin{itemize}

\item \emph{Scripts}: subscripts, superscripts, or both. Detected as \texttt{msub}, \texttt{msup}, and \text{msubsup} nodes.
\item \emph{Accents}: hats, arrows, bars, etc. Detected as \texttt{mover} nodes with one operator (\texttt{mo}) child, and one identifier child.
\item \emph{Functions}: both declarations (e.g., $p(y|x)$) and usages (e.g., $f(2)$). Detected as an identifier followed by an opening parenthesis, a variable number of nodes, and a closing parenthesis.
\end{itemize}

\begin{figure}

\centering
\includegraphics[width=.47\textwidth]{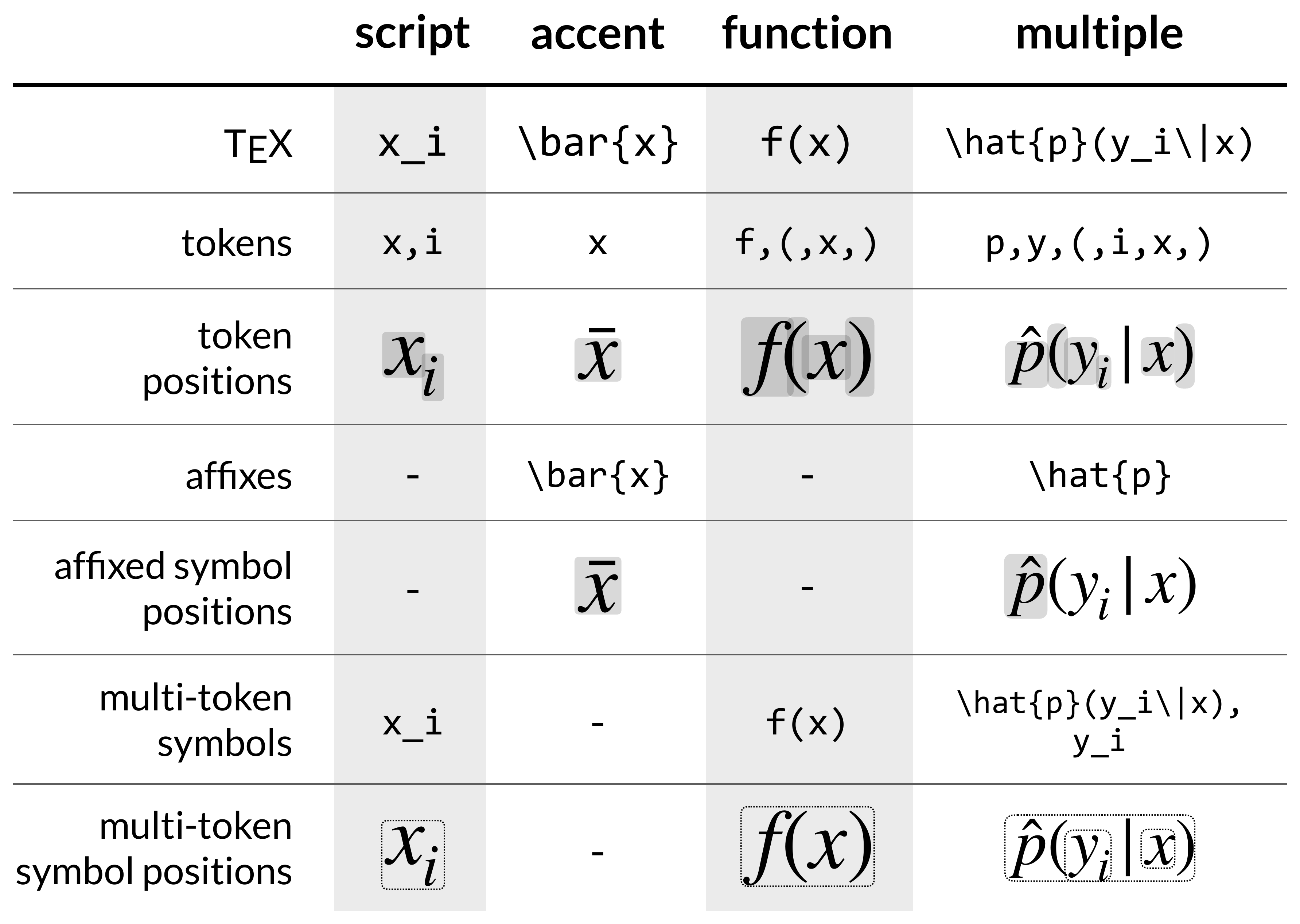}
\caption{Detection of composite symbols. \textmd{Symbols are segmented into 
tokens (i.e., individual characters). The positions of these tokens are found, 
and combined to find the bounding boxes of composite symbols.}}
\label{fig:composite_symbols}
\Description{%
Decorative table with examples of four types of composite symbols in each 
column, and a walkthrough of how their positions are determined based on the 
positions of the sub-symbols they are composed of.

For example, in the first column, the composite symbol ``x-sub-i'' is detected 
by splitting it into tokens (``x'' and ``i''), detecting the positions of both 
of those tokens, and then forming a bounding box for the composite symbol 
``x-sub-i'' that surrounds the boxes for both ``x'' and ``i.''

For symbols with accents (like ``bar-x''), the composite symbol is detected as a 
whole by applying coloring commands around the TeX for both the base and the 
accent.

Also shown is the detection of the positions of all simple and composite symbols 
within the composite symbol for the function ``f(x),'' and a math expression
comprised of all types of composite symbols, ``p-hat of y-sub-i given x''.
}

\end{figure}

The positions of simple symbols are detected using the TeX colorization 
technique described above. The positions of composite symbols are computed as
the minimum bounding box that encapsulates all bounding boxes of simple
symbols the composite symbol is made up of (Figure~\ref{fig:composite_symbols} shows some examples).

On a development set of 12 recent papers from recent proceedings of the
ACL, EMNLP, NeurIPS, and ICML conferences, 
this technique identifies symbols (including both simple and composite symbols) 
with an average precision of 96\% and recall of 88\%. Recall increases to 91\% 
if TeX macros are expanded before processing. For the paper used as a stimulus in the usability 
study, this technique locates symbols with a precision of 98\% and a recall of 
98\% (albeit omitting symbols that appeared in figures).

\subsection{Definition and Usage Recognition}
\label{sec:definitions}

\paragraph{Definition Recognition}
To recognize definitions of nonce words, our implementation has taken three 
approaches. The first approach is to leverage state-of-the-art natural language processing models for definition recognition.
In research parallel to this project, we have developed new models for definition recognition~\cite{ref:kang2020heddex}, 
attaining state-of-the-art results with 73\% precision and 74\% recall on the 
W00~\cite{ref:jin2013mining} dataset. We are continuously improving these models.

A second approach has been to identify abbreviations and expansions with
state-of-the-art models, like those reviewed in Veyseh et 
al.~\cite{ref:veyseh2020acronym}.  These models regularly expand abbreviations 
with an accuracy above 90\%.  A third approach appropriate for prototyping is to 
develop linguistic rules for extracting definitions, like searching for
noun phrases that appear just before symbols, like the word ``encoder'' in the 
passage ``The encoder $E$ is used to\ldots{}''.

All of the above methods yield occasional errors. Section~\ref{sec:futurework} 
envisions techniques for incorporating human input to improve the quality of 
definitions. As the above methods were fine-tuned on incompatible datasets 
without examples of nonce words or TeX symbols, they naturally do not detect 
them well (in our initial tests on the stimulus paper, abbreviation expansion 
with the Schwartz-Hearst algorithm~\cite{ref:schwartz2002simple} yields 
$precision = 55\%$, $recall = 43\%$; term / symbol definition recognition with 
HEDDEx~\cite{ref:kang2020heddex} yields $precision = 24\%$, $recall = 6\%$).  To 
address this gap, we are developing datasets exclusively composed of nonce words 
to train more advanced models.

\paragraph{Identifying usages and defining formulae}
To identify the usages of a nonce word, ScholarPhi extracts sentences from 
papers and associates each nonce word with the sentences it appears in.  
The \texttt{pysbd}~\cite{ref:pysbd} 
sentence boundary detector is applied to the TeX source for the paper; every sentence that 
the nonce word appears within is considered a usage. The positions of sentences within the PDF
are found via the same colorization technique used to detect the positions of 
equations and symbols.

Defining formulae are extracted for symbols by searching for equations in which 
the symbol appeared on the left-hand side of an equation (i.e., to the left of a 
definition operator like ``\texttt{=}'').

\subsection{User Interface}
\label{sec:uiimplementation}

The ScholarPhi user interface is implemented as an overlay atop the Mozilla 
Foundation's open source \emph{pdf.js} PDF reader application~\cite{ref:pdfjs}.  

\subsubsection{Reflowable Definitions and Usages with Math Expressions}
In the ScholarPhi interface, definitions and usages are reflowable; that is, 
their text can wrap. This allows widgets like tooltips and lists of usages to 
have a reduced footprint, because definitions and usages can be rendered with a 
narrower width than the original text.

Widgets need to render math expressions cleanly, because math appears in many 
definitions and usages. Therefore, definitions and usages are rendered from the 
TeX for a sentence, using the KaTeX browser-based formula rendering 
library to transform TeX equations into resizable, reflowable HTML text 
elements.

\subsubsection{Declutter}
A paper is decluttered by applying a semi-opaque SVG mask over every page, and 
then subtracting from that mask rectangular regions that correspond to all 
appearances of a nonce word, and the sentences they belong to.

\subsubsection{Symbol Search}
When a user selects a symbol, the default search bar for ``Control+F'' text search 
is replaced with a navigation widget that lets users cycle through all 
appearances of the symbol. When the user clicks out of the symbol, the
default search widget is restored. Two symbols are considered to be the same
symbol if they were parsed into the same MathML representation by our paper processing
pipeline (see Section~\ref{sec:symbols}). Two symbols that are semantically the same may
have different surface representations in the TeX (e.g., ``\texttt{\{ x\}}'' versus 
``\texttt{x}''). These surface differences typically disappear when the TeX is parsed into MathML.

\subsubsection{Equation Diagrams}
Equation diagrams are implemented as interactive labels and leader lines 
overlaid on the paper. Labels are shown for each symbol and sub-symbol for which a
definition is available. If a symbol appears in the same equation diagram twice,
only one instance of that symbol is labeled. Labels are placed on the top and
bottom boundaries of an equation with a fixed margin between the edges of the equation
and the labels. They are divided
evenly between the top and bottom of the equation, with a preference to assign 
a label to the side of the equation (i.e., top or bottom) where it will be 
closest to the symbol it defines. Labels are spaced horizontally using 
Labella.js~\cite{ref:labella}, which implements an overlap-free spacing 
algorithm introduced by Dwyer et al.~\cite{ref:dwyer2005fast}.
\end{changes}

%% file: 07-evaluation.tex
\section{Usability Study}
\label{sec:evaluation}

We performed a formal remote usability study to ascertain the answers to the 
following  questions: Do the features of ScholarPhi aid readers' ability to 
understand the use of nonce words when reading complex scientific papers?  Do 
readers elect to use the features when given unstructured reading time? How are 
the features  used to support the reading experience?  

In a within-participants design, we compared the full features of ScholarPhi to 
a simplified version of the interface and a standard PDF reader on a series of 
close reading tasks on a machine learning paper.  The quantitative and 
subjective results were strongly in favor of the affordances supplied by 
ScholarPhi over a standard PDF reader, with one exception.

\subsection{Study Design}

\paragraph{Participants} The criterion for inclusion was having previously read a 
machine learning paper. A total of 27 participants were recruited through 
university and company mailing lists.  18 were doctoral students, 5 were 
Master's students, 3 were undergraduate students, and 1 was a professional 
researcher. 13 of the 27
participants identified their discipline as machine learning, and 21 were 
somewhat or very comfortable with reading machine learning papers. Participants 
were compensated with a \$20 (USD) gift certificate.
All study sessions were 1 hour long and held remotely over Zoom, a video conferencing platform; participant interactions were logged and screen activity was captured. Participants opened the application in a private browser window, and were asked to share their screen with the experimenters.

\paragraph{Stimulus paper}  For this study, all participants read 
\emph{Linguistically-informed self-attention for semantic role labeling} (LISA) 
~\cite{ref:strubell2018linguistically}. (Several examples in Section 
\ref{sec:demo} are drawn from this paper.)  \change{This paper was chosen as it 
is widely-read within the field of natural language processing, yet like many 
other papers, it uses some notation inconsistently and does not define all of 
its symbols explicitly.}

\begin{changes}
As our goal was to assess interaction design independently of the performance of 
the algorithms (which are constantly evolving), a clean set of symbol, 
definition, and usage data was curated. Definitions were extracted by hand.  
Symbols, defining formulae, and usages were extracted automatically 
(Sections~\ref{sec:pipeline} and~\ref{sec:definitions}), with a small number of 
manual corrections. We are planning follow-up studies to examine the impact of 
errors on usability.
\end{changes}

\paragraph{Tasks}

Each session ran as follows:
(1) Greeting and consent form. 
(2) Interactive tutorial with all features on a two-page paper~\cite{ref:cohen2016crater}.
(3) Read the abstract of the stimulus paper.
(4) Complete a timed practice question with the full interface.
(5) Complete three timed test questions using each of the three test interfaces (4 minutes each), each followed with a question about confidence and ease of use.
(6) Unstructured reading of the stimulus paper (15-20 minutes).
(7) Questionnaire on background and subjective responses.

In the unstructured reading portion participants were encouraged to make use of 
the tools if they anticipated they would be helpful.  The intention of this 
segment was to observe which aspects of the tool were used when not under time 
pressure.

\paragraph{Interfaces}

Three interfaces were compared within-participants:

\begin{itemize}
    \item ``\plain'' is a basic PDF reader with standard search functionality 
(specifically, being able to find words using ``Control-F'' with a toggle button 
to match case and the ability to highlight all matches).
    \item  ``\sesame'' is a PDF reader with additional declutter functionality.
    \item ``\everything'' is a PDF reader with all ScholarPhi features.
\end{itemize}

\paragraph{Test questions} The three multiple-choice  test questions were each 
intended to assess a different aspect of pain points identified by formative 
studies.  \begin{itemize}
    \item ``\qone'': ``According to Table 1, which model achieves the best recall on WSJ data when GloVE embeddings are used?''
    \item ``\qtwo'': ``Which text corpora is the ConLL-2005 dataset made from? Select all that apply.''
    \item ``\qthree'': ``What does T (upper case) mean in this paper? Select all senses in which T is used.''
\end{itemize}

\paragraph{Assessment measures}

For each of the test questions, we measured the following quantitative metrics:

\begin{itemize}
    \item ``\conf{}'' is a five-point Likert scale variable indicating the participant's self-assessment of the following prompt: ``I am confident I came up with the right answer.'' A score of 5 indicates strong agreement, and a score of 1 indicates strong disagreement.
    \item ``\ease{}''  is a five-point Likert scale variable indicating the participant's self-assessment of the following prompt: ``It was easy to find the answer.'' A score of 5 indicates strong agreement, and a score of 1 indicates strong disagreement.
    \item ``\timeVar{}'' is the number of seconds the participant spent to answer the question. It is measured from when the question first appeared on the participant's screen, to when the participant clicked the next button or the question timer expired (whichever event occurred first).
    \item ``\acc{}'' is a binary variable indicating whether the participant's response was correct. For questions requiring a response with multiple selections, a response was considered correct  if it included  all and only the correct selections. 
    \item ``\area'' is the proportion of the full paper viewed. It is computed as the cumulative total area viewed over the total area in the entire paper. It ranges between values 0 (none of the paper viewed) and 1 (entire paper viewed).
    \item ``\dist'' is a continuous variable measuring the cumulative (normalized) absolute vertical pixel distance---that is, number of document lengths---traversed by a participant.   Normalization controls for different pixel heights across participants' devices. The distance between the top and bottom pixels on each page is set to $1/n_{pages}$ such that the entire paper's total height sums to $1.0$; traversing the length of the paper twice would contribute $2.0$ to the total \dist. 
\end{itemize}

\paragraph{Unstructured reading task measurements}

Measurements in the unstructured reading tasks included frequency of usage of key features 
and subjective feedback.

\paragraph{Assignment} Using a repeated measures factorial design, we assigned each participant to three of nine possible configurations---interface-question pairs---while ensuring that \textit{(i)} each participant observed each interface and each question type exactly once and \textit{(ii)} all nine configurations had the same number of assigned participants.  \change{
Assignment was counterbalanced such that no one interface, question, or interface-question pair was 
experienced more often than others as the first, second, or third task. 9 
participants received interfaces in the order [\plain{} (B), \sesame{} (D), 
\everything{} (S)], 9 in the order [D, S, B], and 9 in the order [S, B, D].
}

\paragraph{Analysis} For each of the quantitative measurements, we fit a generalized linear mixed-effects model (GLMM) with fixed effects for the interface and question factors (and a fixed-effects interaction term). Details can be found in Appendix \ref{sec:appendixa}.

\paragraph{Reduced controls due to remote testing}

Since the study was held remotely, some standard controls could not be employed: 
the size of the screen, the speed of the user's computer (the PDF reader 
appeared to have lag for some participants and not for others), and the 
distraction in the environment (background noise could be heard for many of the 
participants). These differences might account for variation in performance and 
subjective accounts of the experience. Rather than degrading the quality of the 
data, these factors make the study better represent the variation that we 
anticipate readers using this tool would have in their environments.

\subsection{Quantitative Results}
\label{sec:quantitative}

\begin{figure}

\centering
\includegraphics[width=.44\textwidth]{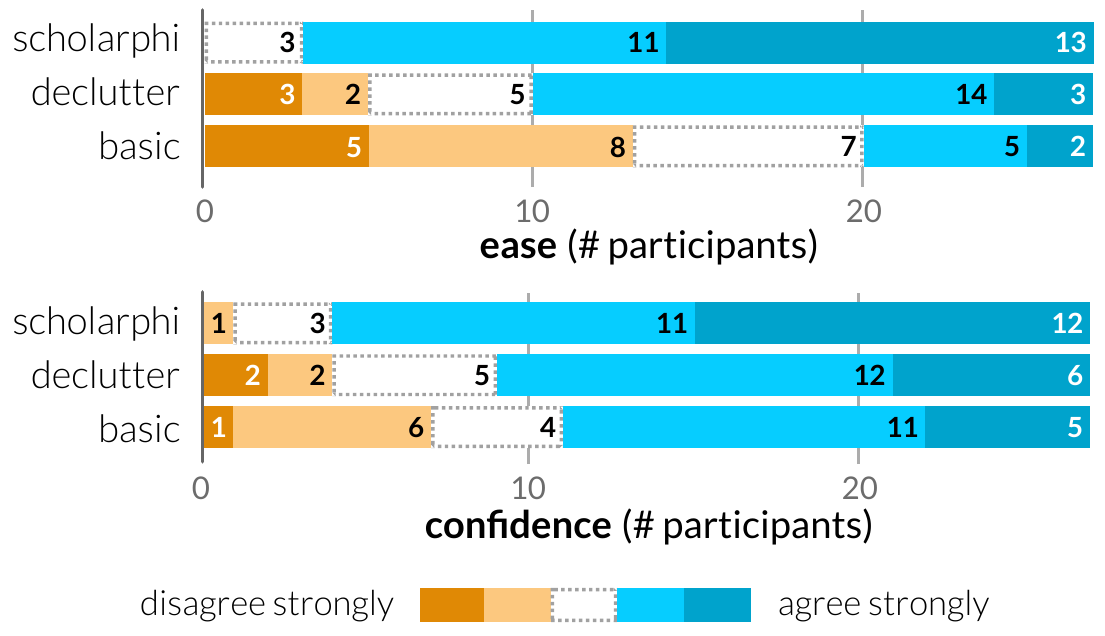}
\caption{Subjective responses for test questions. \textmd{Larger counts of
agreement are preferred. Both measures were reported on an ordinal scale 
with levels ``disagree strongly,'' ``disagree somewhat,'' ``neutral,'' 
``agree somewhat,'' and ``agree strongly.''}}
\label{fig:subjective-measures}
\Description{%
Stacked bar charts.

The first chart is for the ``Ease'' measurement. What follows are counts of 
participant responses to the prompt ``It was easy to find the answer,'' with 
five numbers for each interface: The number of participants who disagreed 
strongly, disagreed somewhat, reported neutral, agreed somewhat, and agreed 
strongly. Data table follows:

ScholarPhi: 0, 0, 3, 11, 13
Declutter: 3, 2, 5, 14, 3
Basic: 5, 8, 7, 5, 2

The second chart is for the ``Confidence'' measurement. What follows are counts 
of participant responses to the prompt ``I am confident I came up with the right 
answer,'' with five numbers for each interface: The number of participants who 
disagreed strongly, disagreed somewhat, reported neutral, agreed somewhat, and 
agreed strongly. Data table follows:

ScholarPhi: 0, 1, 3, 11, 12
Declutter: 2, 2, 5, 12, 6
Basic: 1, 6, 4, 11, 53
}

\end{figure}

\begin{figure}

\centering
\includegraphics[width=.47\textwidth]{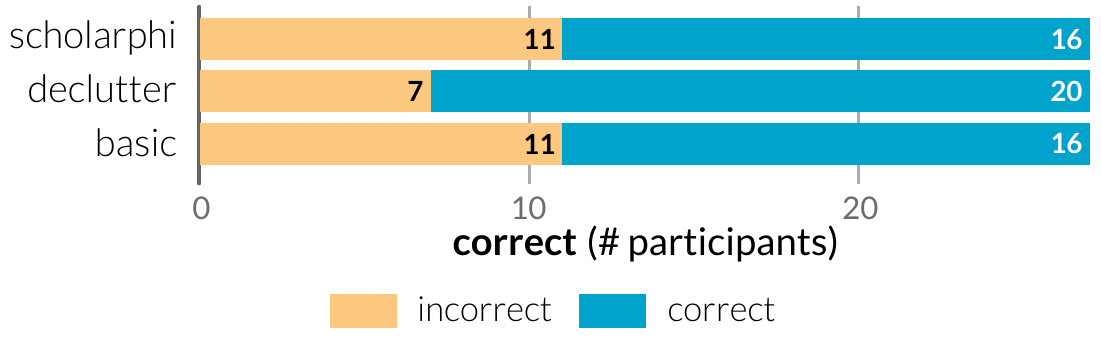}
\caption{Correctness of responses for test questions. \textmd{A lower number of 
incorrect answers is preferred.}}
\label{fig:correctness}
\Description{%
Stacked bar chart, representing the number of correct and incorrect test question responses with each interface.

Data table follows:
Columns: Interface, Number of incorrect responses, Number of correct responses
Rows:
ScholarPhi, 11, 16
Declutter: 7, 20
Basic: 11, 16
}

\end{figure}

\begin{figure}

\centering
\includegraphics[width=.47\textwidth]{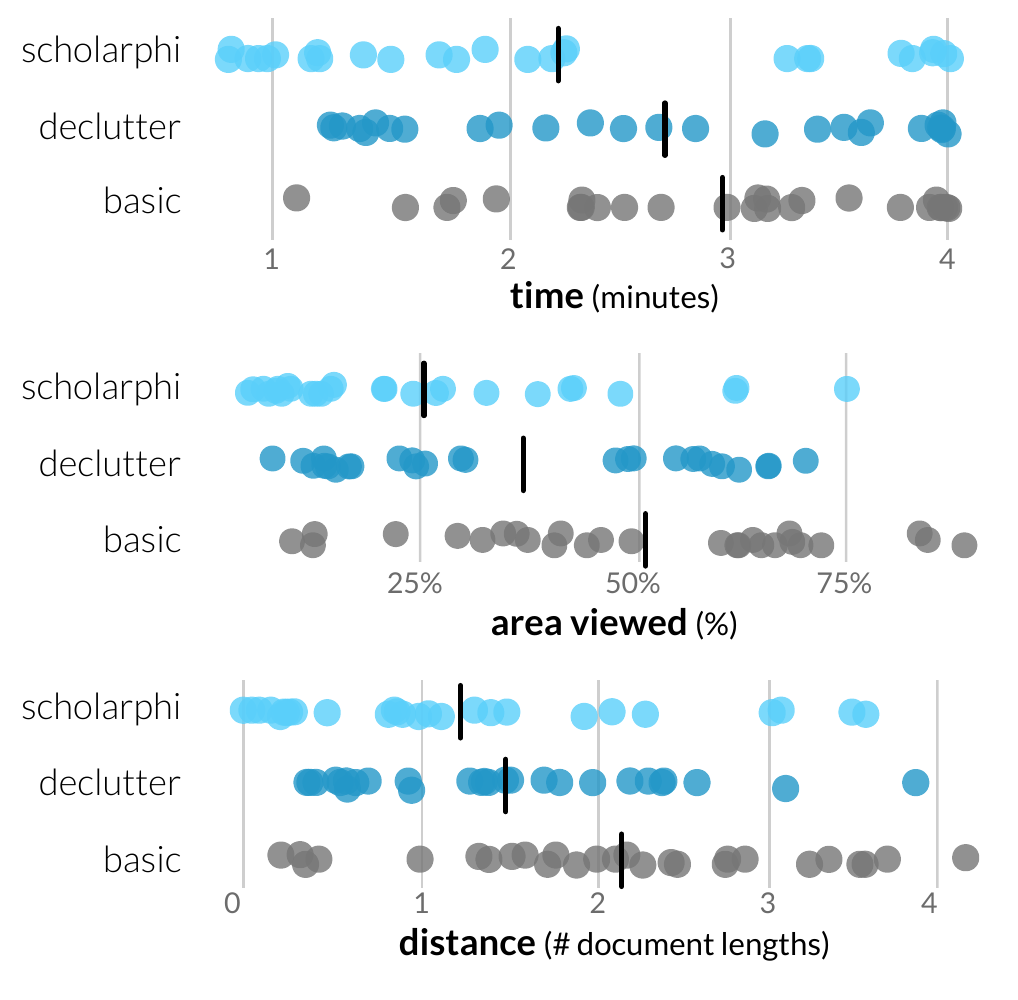}
\caption{Quantitative results for test questions. \textmd{The vertical bars 
indicate the mean. Lower values are preferred.}}
\label{fig:continuous-measures}
\Description{%
Jittered dot plots of interaction log data from the usability study. Each plot shows the same measured variable along the x-axis against three interface conditions in the y-axis: using ScholarPhi with all the features, using only the declutter feature, and using a baseline basic PDF reader. Three plots:

Plot 1. Time taken to answer question in minutes.

ScholarPhi: Mean: 2.2, Standard deviation: 1.19
Declutter: Mean: 2.7, Standard deviation: 1.08
Basic: Mean: 3.0, Standard deviation:0.88

Plot 2. Percentage of total document area viewed.

ScholarPhi: Mean: 25, Standard deviation: 19.1
Declutter: Mean: 37, Standard deviation: 20.9
Basic: Mean: 51, Standard deviation: 22.4

Plot 3. Distance traveled in the document in number of document lengths.

ScholarPhi: Mean: 1.2, Standard deviation: 1.08
Declutter: Mean: 1.5, Standard deviation: 0.91
Basic: Mean: 2.1, Standard deviation: 1.10
}

\end{figure}

Figures~\ref{fig:subjective-measures},~\ref{fig:correctness}, and~\ref{fig:continuous-measures}
summarize how the measures on the test questions vary across the three 
interfaces.  We report results from two-sided tests for pairwise 
differences in mean effects between interfaces in Table~\ref{tab:pairwise-contrasts}. These results indicate which 
trends in the figures are statistically 
significant at the $\alpha = 0.05$ level.

In terms of the subjective scores, we observed that \everything{} outperformed \plain{} on \ease{} and \conf{}, and \sesame{} on \ease{}. \sesame{} also reported higher \ease{} than \plain{},
but not higher \conf{}. \everything{} reported higher \conf{} than \sesame{}, but this result was not significant at $\alpha = 0.05$.

\everything{} outperformed the other interfaces in 
terms of time required to answer the test questions (\timeVar) (\sesame{} and 
\plain{} were not significantly different). \change{No statistically significant 
differences were observed in the number of participants who answered questions 
correctly among the three interfaces}.

Finally, we observed that participants traversed less screen \dist{} and viewed 
less \area{} of the paper under \everything{} and \sesame{} compared to \plain; 
\everything{} outperformed \sesame{} on \area{} but did not significantly outperform 
\sesame{} on \dist. Overall, these results suggest that even the lighter-weight 
version of the tool, with the declutter overlay alone, yields benefits over the 
standard PDF reader, but the full set of features in \everything{} is especially 
beneficial.

Upon further inspection of the results on \acc, we found the performance of 
participants on a particular question yielded the reason for \everything{} 
performing similarly to \plain{} (with \sesame{} yielding slightly higher results):  
participants performed better on both the \qone{} and \qtwo{} questions using \everything, but 
performed very poorly on the \qthree{} question with this interface.   Recall from Section~\ref{sec:design-motivations} (M1) that the stimulus paper uses 
the symbol $T$ inconsistently and also does not define all senses of this 
symbol.  We found that participants almost always answered this question 
incorrectly using \everything{} because the definitions list did not show all of the 
usages, and the participants had the expectation that the definitions list showed all 
senses of the symbol.  This highlights an important potential drawback of a 
tool like ScholarPhi: it can mislead if it implies incorrect information.

\input{tables/pairwise-contrasts}

\subsection{Qualitative Results}

When describing qualitative results, we refer to participants as ``readers,'' 
and to individual readers with pseudonyms P1--27. 

\subsubsection{Subjective Impressions}

Subjective responses were obtained both from oral comments during the study and 
from open-ended questions in the final questionnaire. Readers' impressions of 
ScholarPhi were overwhelmingly positive. Readers were
enthusiastic about the support that ScholarPhi provided for the reading task.
They described the tool as ``cool'' (P8), ``very cool'', (P13), ``super cool'' 
(P12), and ``amazing'' (P4, P16, P19). Eight of the 27 responses to the 
open-ended questionnaire forms contained exclamation marks conveying reader 
excitement for the tool. Several readers commented on the polish of the 
prototype (P7, P24).

Readers described three supporting roles they envisioned ScholarPhi playing 
during
reading tasks. First, they believed ScholarPhi would help them
maintain ``reading flow'' (P16, P27). In the words of one reader, ScholarPhi 
helped them ``focus on the aspects of the paper that interested me, and not 
waste time on other stuff'' like reminding themselves of definitions (P4).
The features provided timely reminders  (P10, P21, P26), and eliminated the need 
to traverse ``back and forth'' within the paper (P11).

Second, ScholarPhi helped them ``check their understanding'' of the meanings of 
nonce words (P16) and the passages of text they appeared in (P20). Third, 
readers believed ScholarPhi could help them understand papers that they 
otherwise would not have had the vocabulary to read easily (P4, P23), in effect 
``lowering the barrier'' to reading papers in fields outside of one's expertise.

\paragraph{Anticipated usage}

To determine which of ScholarPhi's features would be of greatest interest to 
readers in the future, and hence which features should be developed further, 
readers were asked to report how often they expected they would use each feature 
if it was available in the software they used to read papers.  \change{Expected 
frequency was reported on a five-point ordinal scale (``Never,'' ``Rarely,'' 
``Sometimes,'' ``Often,'' ``Always,'' and ``Unsure'').}

Readers expected they would use most features often. They envisioned using 
multiple features very frequently, including definition tooltips for symbols (16 
``always''; 8 ``often''), definition tooltips for terms (15; 9), and equation 
diagrams (17; 6). The features of decluttering for symbols (5; 13) and terms (2; 
15), and the priming glossary (8; 6), were envisioned as being used less 
frequently.  While a reader could indicate they ``never'' saw themselves using a 
feature, not a single reader selected this option for any feature.

\subsubsection{Use of Features}

To identify strengths in the design and opportunities for improvement, usage 
logs were inspected, and participant feedback on individual features was 
reviewed. All readers except for one (96\%) used at least one of ScholarPhi's 
features during the unstructured reading time.  Analysis of the aforementioned 
data led to the following observations about feature design:

\paragraph{Definition tooltips}

For most readers, tooltips were ScholarPhi's most essential feature.  During 
unstructured reading time, readers used definition tooltips more than any other 
feature. All but three readers opened at least one tooltip for a symbol, and all 
but one reader opened at least one tooltip for a term. When readers used 
tooltips they used them often. Readers opened tooltips for symbols a median of 
10 times ($\sigma = 13.8$, $max = 54$), and for terms a median of 5 times 
($\sigma = 3.6$, $max = 14$).

Tooltips served two purposes for readers. The first was the purpose they were 
designed for: to provide access to definitions of nonce words that appeared 
elsewhere in the paper (P10). A second purpose was to help a reader check 
whether the passage the reader was consulting was indeed the definition of a 
nonce word, which could help a reader make sure they were not missing other 
information of interest about the nonce word (P2).

\paragraph{Declutter}

In contrast to tooltips, which were unanimously appreciated, the declutter 
feature saw disagreement. Some readers valued the feature, and others did not.

On the whole, readers' behaviors suggests that most readers expected declutter 
to be useful for finding answers to questions in a paper: all participants 
activated declutter at least once in the test task where they used an interface 
with only the declutter feature enabled. Several readers explicitly told us they 
believed declutter could be useful for finding information about nonce words 
(P6, P11, P15, P23, P26). Readers reported that the feature made the paper look 
``less cluttered,'' and that it could help them feel ``less overwhelmed'' by the 
text in the paper (P27).

Other readers indicated gaps in the design.  Some readers did not understand the 
point of the feature (P25), or thought it provided little value over the 
definition tooltips (P22). Others felt that the standard ``Control-F'' search 
provided a more efficient interface for searching a paper than scrolling through 
a paper with declutter (P2). An additional gap of the feature is that, unlike 
``Control-F'' search, declutter cannot be invoked unless the nonce word of 
interest is already in view. One reader believed this would be frustrating in 
the scenario where they temporarily deactivated declutter in order to read the 
low-lighted text and then wished to resume declutter for the same nonce word as 
before (P14).

\paragraph{Lists of usages}

Nearly all (20 of 27) readers opened a list of definitions, defining formulae, 
or usages during the unstructured reading task. 18 readers opened a list of 
definitions, 3 opened a list of defining formulae, and 10 opened a list of 
usages. Some readers used the lists heavily. For instance, one participant 
opened the lists of definitions and usages eight times each (P4).

Readers reported that they used the list of usages to develop an understanding of the 
purpose of the paper (P9) and gather context to check their understanding of a 
term (P16). One reader described the list of usages as a ``guide''
to support non-linear reading (P27). They navigated the paper by iteratively
selecting nonce words, reviewing usages, jumping to a usage, and then looking 
for other nonce words of interest in the passage they jumped to. This reader 
believed the list helped them answer questions as they came up, rather than 
waiting them to be resolved in a later passage.

\paragraph{Equation diagrams}

More readers expected they would ``always'' use equation diagrams for future 
readings than any other feature. Almost all (21 of 27) readers
opened an equation diagram during the unstructured reading task. Most readers 
opened multiple, with the median reader opening 3 ($\sigma = 4.3$, $max = 14$).

The primary use of equation diagrams was to understand the symbols in an 
equation without attending to the surrounding text (P1, P6, P11, P13, P14, P21, 
P24). Diagrams were seen as particularly useful when an equation was long (P24) 
or complex (P11). One of the equations, for instance, consisted of four lines of 
notation with a total of fourteen symbols for which definitions were available, 
and many others for which definitions were not. Readers were regularly observed 
pausing to study this equation with the diagram open.

Beyond the primary use of describing symbols, one reader described diagrams as 
supporting a new way of navigating the text. This reader skimmed the technical 
section of the paper by opening the diagrams one-by-one, familiarizing 
themselves with the section by reading the equations rather than the prose (P7).

\paragraph{Priming glossary}

The priming glossary was the least-used feature during the unstructured reading 
task. A few readers (6 of 27) were observed consulting the priming glossary for 
a nontrivial amount of time, defined in our protocol to be 10 or more seconds.  

Although readers infrequently consulted the priming glossary, a few readers 
believed it would be useful under certain circumstances. Some readers believed 
the glossary could help them orient to the terminology used in a paper before 
reading it (P13, P16). In line with this claim, one reader spent 2 minutes (P16) 
and another spent 5 minutes (P1) carefully studying the glossary at the 
beginning of the unstructured reading time. Second, readers indicated an 
expectation that the glossary would provide more thorough definitions of nonce 
words than the tooltips. Several readers appeared to visit the glossary as a 
fallback when the definition tooltip did not contain the information they sought 
(P3, P12, P14, P22).

\paragraph{Use of features in concert}

While ScholarPhi's features were often used in isolation, we also observed on 
several equations readers using several disparate features in rapid succession.  
For example, P6 clicked an equation to reveal a diagram, selected one of the 
symbols in the diagram, opened the list of definitions for the symbol, and then 
clicked on a link that took them to one of those definitions. Several readers 
chained interactions across multiple of ScholarPhi's features in a similar way 
(P6, P8, P13, P19). 

%% file: tables/pairwise-contrasts.tex
\begin{table*}
\centering
\begin{tabular}{rcccccc}
\toprule &
\if 0
\multicolumn{2}{c}{
  \begin{tabular}[x]{@{}c@{}}\everything{}  vs. \sesame{} \end{tabular}
} &
\multicolumn{2}{c}{
  \begin{tabular}[x]{@{}c@{}}\sesame{} vs. \plain{} \end{tabular}
} &
\multicolumn{2}{c}{
  \begin{tabular}[x]{@{}c@{}}\everything{} vs. \plain{} \end{tabular}
}
\\
&
\fi
$\hat{y}_S - \hat{y}_D$ &%
$p$ &%
$\hat{y}_D - \hat{y}_B$ &%
$p$ &%
$\hat{y}_S - \hat{y}_B$ &%
$p$ \\

\midrule              

\conf{} (1--5) &%
0.59&%
0.094&%
0.19&%
0.785&%
0.78&%
\textbf{0.020} \\

\ease{} (1--5)&%
0.93&%
\textbf{0.005}&%
0.78&%
\textbf{0.020}&%
1.70&%
\textbf{\textless{}0.0001} \\

\timeVar{} (seconds)&%
-27.6&%
\textbf{0.015}&%
-16.8&%
0.218&%
-45.4&%
\textbf{0.0001} \\

\acc{}&%
-15\%&%
0.393&%
15\%&%
0.393&%
0\%&%
1.000 \\

\dist{} (\# doc lengths)&%
-0.24&%
0.572&%
-0.66&%
\textbf{0.023}&%
-0.90&%
\textbf{0.001} \\

\area{}&%
-11\%&%
\textbf{0.047}&%
-14\%&%
\textbf{0.009}&%
-25\%&%
\textbf{\textless{}0.0001} \\

\bottomrule
\vspace{0.2ex}
\end{tabular}

\caption{
Two-sided tests for pairwise differences in mean effects between interfaces. \textmd{This table reports
$\hat{y}_i - \hat{y}_j$ and Holm-Bonferroni-corrected $p$-values~\cite{holm79}, 
where $\hat{y}$ is the estimated mean of $y$ under the GLMM, and $i,j$ 
correspond to interface options --- $B=\plain$, $D=\sesame$, $S=\everything$.  
For example, in the cell for (\timeVar{}, $\hat{y}_S - \hat{y}_B$), we can 
interpret the result as \everything{} is associated with tasks completed in 45.4 
fewer seconds in \timeVar{} than \plain{}, on average.  \acc{} and \area{} differences are 
reported as absolute, not relative, percentage point differences. Statistically 
significant $p$-values are bolded.  Further details about this analysis appear in 
Appendix \ref{sec:appendixa}.}
}
\label{tab:pairwise-contrasts}
\Description{%
Data table of pairwise comparisons of key usability metrics between pairs of 
interfaces. Statistically significant pairwise differences are reported in the 
text.
}
\end{table*}

%% file: 08-agenda.tex
\section{Discussion and Future Work}

\subsection{Summary of Results}

The outcomes of the usability study produced the following answers to the research questions:

\paragraph{Do the features of ScholarPhi aid readers' ability to understand the use of nonce words when reading complex scientific papers?}

Yes. When asked to answer questions requiring understanding of nonce words,
readers answered questions significantly more quickly with ScholarPhi than with 
a baseline PDF reader, while viewing significantly less of the paper.

\paragraph{Do readers elect to use the features when given unstructured reading time?}

Yes. 96\% of readers used ScholarPhi's features at least once during 15 minutes 
of unstructured reading time. Tooltips were the most frequently used feature: 
readers  opened a median of 10 tooltips for symbols, and 5 for terms. Equation 
diagrams were opened a median of 3 times. Almost all participants opened a list 
of definitions, defining formulae, or usages at least once.

\paragraph{How are the features used to support the reading experience?}

On the whole, readers used the features for the reasons expected: they referred 
to tooltips to remind themselves of forgotten definitions, activated declutter 
to find information about nonce words within a less cluttered view of the paper, and 
opened equation diagrams to view the definitions of many symbols at once.  
Readers also used the tools to support the reading experience in
unconventional ways, for instance using the list of usages as a ``guide'' to 
support a non-linear, curiosity-driven reading, and skimming a section by 
jumping from one equation diagram to the next.

\subsection{Limitations}

A major limitation of the usability study is its  focus
on a single paper, where performance was measured for only three tasks. Papers 
vary widely in clarity and readability.  To improve generalizability of the 
study, the paper was selected to be a widely-read scientific paper
exhibiting some of the very problems the system was seeking to address.  
Furthermore, the three tasks were chosen to require an understanding of 
different types of nonce words: terms referring to datasets, baselines, and symbols.  
In the future, we will continue to evaluate ScholarPhi on a variety of research 
papers, as has been done to date through the iterative design process for the 
tool.

A second limitation, that pertains to the tool's suitability for
supporting unstructured reading, is that readers in the study only used the tool 
for 15--20 minutes, and may have not had enough time to discover limitations 
that would preclude them using the tool in the future. Observations from our 
pilot studies have suggested that readers continue to find aspects of the tool 
useful after 20 minutes of reading, but longitudinal studies are necessary to
better assess how readers would employ ScholarPhi in day-to-day use. 

\subsection{Future Work}
\label{sec:futurework}

The study of ScholarPhi has revealed three opportunities for future research to 
advance the potential of intelligent reading interfaces to aid in the authoring 
and reading of scientific papers.

\paragraph{Connecting Readers to Definitions Beyond the Paper} \change{The 
larger vision of ScholarPhi is to help scientists more easily read papers by 
linking relevant information to its location of use.  This includes providing 
links to the contents of cited papers, and providing definitions going beyond 
nonce words to terms defined externally to the paper.  Indeed, readers in the 
formative study, pilot studies, and usability study all asked for the ability to 
look for definitions of terms that resided outside of a paper. Future work will 
incorporate this information into the ScholarPhi reader. }

\paragraph{Co-development of Reading Interfaces and Machine Learning Models}

\change{Machine learning models are imperfect; our 
own recent research~\cite{ref:kang2020heddex}} shows that the state-of-the-art 
algorithms for definition detection currently have a problem of recall when it 
comes to detecting definitions in scientific papers. 
Researchers in human-computer interaction have explored how users interact with 
imperfect AI algorithms~\cite{ref:yin2019understanding,ref:kocielnik2019will}. ScholarPhi may benefit from an analogous thread of research which explores how models for augmenting texts with interactive affordances can convey uncertainty.  

\change{ 
Definition quality could also be improved by incorporating human input.  
Annotation tools could let authors  explicitly define nonce 
words and then refer to them unambiguously. 
Furthermore, readers could be asked to improve 
definitions by selecting helpful definitions from among a set of alternatives, or 
directly editing the definitions shown in tooltips and equation diagrams.}

\paragraph{ScholarPhi for Writing Scientific Papers}
A dual of ScholarPhi  could support the task of {\em writing} clear scientific papers.
Such a tool could indicate to an author when they left a nonce word undefined, 
when they used the same symbol to mean two different things (as is often the 
case for symbols like ``$k$''), and to know when they are using multiple nonce words
to refer to the same idea. The same paper processing technologies that can 
detect definitions and relate two nonce words to each other could suit writing 
just as well as reading. As we saw in the development of ScholarPhi, the design 
exploration of augmented writing interfaces likely needs to begin with careful 
observations of writers to understand how lightweight, non-intrusive
features can support the writing task without distracting authors.

%% file: 09-conclusion.tex
\section{Conclusion}

The ScholarPhi system was designed to help readers 
concentrate on the cognitively demanding task of reading scientific papers by 
providing them efficient access to definitions of nonce words. The iterative design
of the system revealed that systems like ScholarPhi need to tailor definitions 
to the passage where a reader seeks an understanding of a nonce word, provide scent, 
and avoid distracting readers from their reading. A usability study with 27 
researchers showed that when using ScholarPhi versus a standard PDF reader, they 
could answer questions that required an understanding of nonce words in less time, 
viewing less of the paper. Readers could see using ScholarPhi's 
definition tooltips and equation diagrams ``often'' or ``always'' if they were 
available in their reading interface. These strong empirical results suggest 
that researchers are eager and ready for tools like ScholarPhi that support the 
reading task by providing just-in-time, position-sensitive definitions
of nonce words when and where they need them.

%% file: 10-appendix.tex
\appendix

\section{Appendix}

\subsection{Statistical Analysis}
\label{sec:appendixa}

\subsubsection{Modeling Mixed-Effects in Repeated Measures Studies}

For the analysis in Section~\ref{sec:evaluation}, we used the generalized linear 
mixed-effects model (GLMM). GLMMs are often used to analyze repeated measures, 
in which the same subject contributes multiple (potentially correlated) 
measurements~\cite{Lindstrom1990NonlinearME}. They have been used to analyze
measurements from studies in medicine~\cite{Cnaan1997UsingTG}, the behavioral 
sciences~\cite{cudeck-1996-mixed-models-repeated-measures-behavioral}, and 
human-computer interaction~\cite{Hearst2020AnEO}.

\subsubsection{F-Tests for Significant Effect of Interface}

For each of the quantitative measurements ($y$), we fit a GLMM with fixed 
effects $\beta$ for the interface ($x_1$) and question ($x_2$) factors (and a 
fixed-effects interaction term). The models were fit using the \textsc{lme4} 
package in R~\cite{Bates2014FittingLM}. More precisely, we fit the following 
GLMM:

\begin{equation}
    g(E[y]) = \beta_0 + \gamma_j + \beta_1 x_1 + \beta_2 x_2 + \beta_3 x_1 x_2,
\end{equation}

where $g$ is the link function, and the random intercepts $\gamma_j \sim 
\mathcal{N}(0, \sigma^2_{\gamma})$ capture individual variation of each 
participant $j$. For \ease{}, \conf{}, \timeVar{}, \dist{}, and \area{}, we used
the identity link $g(z) = z$.  For \acc{}, which we treated as a Bernoulli 
variable, we used the logit link $g(p) = \log (p / (1-p))$.

Using the \textsc{lmerTest} R package~\cite{Kuznetsova2017lmerTestPT}, we 
conducted \emph{F}-tests for differences in fixed-effect estimates between each 
interface option, repeated for each $y$.\footnote{The \emph{F}-test is not 
applicable when $y \sim$ Bernoulli, so we performed the similar, but slightly 
more conservative, likelihood ratio test for $y = \acc$ 
\cite{Kuznetsova2017lmerTestPT}.} We performed Holm-Bonferroni \cite{holm79} 
correction on the $p$-values using the \textsc{p.adjust} R package. We found
significance for \acc{} $(p=.047)$, \ease{} $(p<.001)$, \conf{} $(p=.040)$, 
\timeVar{} $(p<.001)$, \dist{} $(p=.005)$, \area{} $(p<.001)$---even while 
controlling for question and participant-specific effects.  That is to say, for 
these metrics, the \emph{F}-test has identified that the choice of interface 
(\plain{}, \sesame{}, or \everything) is a significant factor.  Note that the 
\emph{F}-test does not assess \emph{which} of these interfaces is more or less 
impactful on the metric.  

\subsubsection{Tests for Pairwise Differences in Mean Effects between Interfaces}

We conducted a post-hoc analysis to quantify the pairwise differences in mean 
effects between interfaces on $y$ under the GLMM (and controlling for question).  
Two-sided $t$-tests for pairwise comparisons were computed using the 
\textsc{emmeans} R package, yielding the results shown in 
Table~\ref{tab:pairwise-contrasts}.

Because the GLMM for $y=\acc$ was fit using a logit link, direct testing of 
pairwise comparisons $\hat{y}_i - \hat{y}_j = \hat{Pr}(\acc = 1 \vert i) - 
\hat{Pr}(\acc = 0 \vert j)$ was not possible.  We used the \textit{transform} 
option in \textsc{emmeans} to perform the tests on the log-odds $\log Pr(\acc = 
1) / Pr(\acc = 0)$ scale, which are linear under the GLMM, before applying the 
inverse-link $g^{-1}$ transformation to return to the probability $Pr(\acc = 1)$ 
scale.  This yielded the estimated (absolute) differences in reported in 
Table~\ref{tab:pairwise-contrasts}.

\subsubsection{Ordinal Regression for Likert-Scale Variables}

As \ease{} and \conf{} were measured on a 5-point Likert scale, a linear GLMM 
estimated means was seen as potentially ill-suited for analysis, especially if 
\ease{} and \conf{} are not sufficiently normally distributed. We additionally 
performed likelihood ratio tests after fitting analogous cumulative link 
mixed-effects models (CLMM) provided in the \textsc{ordinal} R package 
\cite{Christensen2018CumulativeLM}. Likelihood ratio tests, which are similar to 
\emph{F}-tests but more conservative, yielded similar $p$-values---\ease{} ($p < 
.001$) and \conf{} ($p = 0.045$)---and resulted in the same conclusions as those 
when using the GLMM.  Since pairwise comparisons are not available through 
\textsc{emmeans} (or other libraries) for CLMMs, we opted to use the GLMM model 
for \ease{} and \conf{} to enable subsequent analysis for Table 
\ref{tab:pairwise-contrasts}.